# Control Complexity in Bucklin and Fallback Voting[*]

Gábor Erdélyi
School of Economic Disciplines
University of Siegen
57076 Siegen, Germany

Michael Fellows
Parameterized Complexity Research Unit
Charles Darwin University
Darwin, NT 0909 Australia

Jörg Rothe  and  Lena Schend
Institut für Informatik
Heinrich-Heine-Universität Düsseldorf
40225 Düsseldorf, Germany

August 21, 2012

## Abstract

Electoral control models ways of changing the outcome of an election via such actions as adding/deleting/partitioning either candidates or voters. These actions modify an election's participation structure and aim at either making a favorite candidate win ("constructive control") or preventing a despised candidate from winning ("destructive control"). To protect elections from such control attempts, computational complexity has been investigated with an eye to showing that electoral control, though not impossible, is at least not always computationally easy, unless P equals NP. Such hardness results are termed *resistance*, and it has been a long-running project of research in this area to classify the major voting systems in terms of their resistance properties.

We show that fallback voting, an election system proposed by Brams and Sanver [BS09] to combine Bucklin with approval voting, is resistant to each of the common types of control except to destructive control by either adding or deleting voters. Thus fallback voting displays the broadest control resistance currently known to hold among natural election systems with a polynomial-time winner problem. We also study the control complexity of Bucklin voting and show that it performs at least almost as well as fallback voting in terms of control resistance.

---

[*]Preliminary versions of this paper appeared in the proceedings of *Computing: the 16th Australasian Theory Symposium* [ER10], in the proceedings of the *Tenth International Conference on Autonomous Agents and Multiagent Systems* [EPR11], and in the proceedings of the *Third International Workshop on Computational Social Choice* [EF10a]. This technical report merges, extends, and supersedes the technical reports [EPR10a, EPR10b, EF10b]. This work was supported in part by DFG grants RO 1202/11-1, RO 1202/12-1 (within the European Science Foundation's EURO-CORES program LogICCC), and RO 1202/15-1, by the National Research Foundation (Singapore) under grant NRF-RF 2009-08, and by SFF grant "Cooperative Normsetting" of HHU Düsseldorf. This work was done in part while the first author was affiliated to Heinrich-Heine-Universität Düsseldorf and to Nanyang Technological University, Singapore, and while he was visiting Universität Trier, while the third author was visiting the University of Rochester, and while the first and third authors were visiting NICTA, Sydney, and the University of Newcastle in Australia. Authors' URLs: www.uni-siegen.de/fb5/dt/team/erdelyi, mrfellows.net, ccc.cs.uni-duesseldorf.de/~rothe, ccc.cs.uni-duesseldorf.de/~schend.

As Bucklin voting is a special case of fallback voting, each resistance shown for Bucklin voting strengthens the corresponding resistance for fallback voting.

Such worst-case complexity analysis is at best an indication of security against control attempts, rather than a proof. In practice, the difficulty of control will depend on the structure of typical instances. Parameterized complexity is a way of accounting for such structure in studying computational complexity. We investigate the parameterized control complexity of Bucklin and fallback voting, according to several parameters that are often likely to be small for typical instances. Our results, though still in the worst-case complexity model, can be interpreted as significant strengthenings of the resistance demonstrations based on NP-hardness.

# 1   Introduction

Since the seminal paper of Bartholdi et al. [BTT92], the computational complexity of *electoral control* has been studied for a variety of voting systems. Unlike *manipulation* [BTT89, BO91, CSL07], which models attempts of strategic voters to influence the outcome of an election by casting insincere votes, *control* models attempts of an external actor, the "chair," to tamper with an election's participation structure so as to alter its outcome via such actions as adding/deleting/partitioning either candidates or voters. A third way of tampering with the outcome of elections is *bribery* [FHH09, FHHR09a], which shares with manipulation the feature that votes are being changed, and with control the aspect that an external actor tries to change the outcome of an election. (We do not here investigate resistance to bribery.) Faliszewski et al. [FHH10, FHHR09b] and Conitzer [Con10] comprehensively survey known complexity results for control, manipulation, and bribery for various voting systems; Faliszewski and Procaccia [FP10] do so with a focus on manipulation; and Baumeister et al. [BEH$^+$10] do so with a particular emphasis on approval voting and its variants.

Elections have been used for preference aggregation not only in the context of politics and human societies, but also in artificial intelligence, especially in multiagent systems, and other applied settings in computer science; see, e.g., [ER97, GMHS99, DKNS01]. In general, information increasingly arises from multiple perspectives, and must be collated—one way of thinking about election systems that points to the rich range of applications. The investigation of the computational properties of voting systems is thus well-motivated and the robustness and resistance to manipulation, control, and bribery in voting is of fundamental interest, but how can this be investigated and such desirable properties be established, evidenced, or indicated?

In their path-breaking papers Bartholdi, Tovey, and Trick [BTT89, BTT92] proposed to employ computational complexity to explore this issue: If the problem of deciding whether an election can be tampered with in the manipulation or control scenario at hand is NP-hard, then this is evidence that the system is intrinsically difficult to manipulate or to control in this scenario, because of the computational intractability of mounting an attack. In this perspective, a central quest that has emerged in the past two decades of research is to find natural voting systems with polynomial-time winner determination that are computationally resistant to as many of the common 22 control types as possible, where *resistance* means that the corresponding control problem is NP-hard. Each control type is either *constructive* (the chair of the election seeks to insure that some favored candidate wins) or *destructive* (the chair seeks to insure that some despised candidate does not win).

We study the control complexity of fallback voting, an election system introduced by Brams and Sanver [BS09] as a way of combining Bucklin and approval voting. We prove that fallback voting is resistant to each of the common types of control except two (namely, it is not resistant to destructive control by either adding or deleting voters), and we show that it is vulnerable (i.e., the corresponding control problem is in P) to these two control types. With these 20 control resistances, fallback voting displays the broadest control resistance currently known to hold among natural election systems with a polynomial-time winner problem. In particular, fallback voting is fully resistant to constructive control and it is fully resistant to candidate control.

As the two control types for which fallback voting is vulnerable are *destructive* types and as destructive control intuitively can be seen as less important than constructive control, one may now view the original research project that was started by Bartholdi et al. [BTT92] two decades ago as "essentially completed," and this is one of the main contributions of this paper. That is not to say that, among natural voting systems with polynomial-time winner determination, fallback voting were the one and only system with the strongest or broadest control resistance. Indeed, shortly after our results on fallback voting were made public in the predecessor [EFPR11] (dated March 11, 2011) of this technical report, Menton [Men12] reported analogous results for normalized range voting (the version of his technical report that establishes a matching number of resistances is dated April 25, 2011). It is very well conceivable that also other voting systems have the same resistances and vulnerabilities as fallback voting, and there might even be such a system that in addition is resistant to the two types of destructive control where fallback voting lacks resistance. By "essentially completed" we merely mean that fallback voting is the first natural voting system with polynomial-time winner determination shown to display such an almost complete control resistance.

Fallback voting is a hybrid system combining approval and Bucklin voting, and it is clear that each of these two constituent "pure" systems are certainly more natural than their hybrid. Therefore, we also study the control complexity of Bucklin voting itself. While many important voting systems—including plurality, Condorcet, Copeland, maximin, and approval voting—have already been investigated with respect to electoral control (see the references in "Related Work" below), Bucklin voting is one of the few central voting systems for which a thorough study of the control complexity has been missing to date. We show that Bucklin voting has no more than one control resistance fewer than fallback voting (namely, possibly, regarding destructive control to partition of voters in the tie-handling model TP, see Section 2.2 for the definition). In particular, Bucklin voting is also fully resistant to constructive control and fully resistant to candidate control. Since Bucklin voting is a special case of fallback voting, each resistance result for Bucklin voting strengthens the corresponding resistance result for fallback voting.

Now that the program of investigating electoral control resistance issues in terms of NP-hardness is "essentially completed" with the results in this paper, it is appropriate to revisit the roots of the program. In particular, what do resistance results based on NP-hardness results actually mean in practice, that is, for "typical" elections?

We study the parameterized complexity of some of the key control problems in natural parameterizations. Parameterized complexity allows a more fine-grained deployment of computational complexity, having the ability to model more closely "typical input structure."[1] In particular, we obtain W[2]-hardness results for problems related to adding/deleting either candidates or voters, parameterized by the number of candidates/voters that have been added/deleted.

---

[1] We stress that "typical input structure" is different from (and should not be confused with) the "typical elections" used in, e.g., the experimental results conducted by Rothe and Schend [RS12a, RS12b] for Bucklin, fallback, and plurality voting. While the former refers to typical parameters in the parameterized control problems we study, the latter refers to generating typical elections according to some distribution model.

**Related Work:** The study of electoral control was initiated by Bartholdi et al. [BTT92], who introduced a number of constructive control types and investigated plurality and Condorcet voting in this regard. The common types of destructive control were proposed by Hemaspaandra, Hemaspaandra, and Rothe [HHR07], who studied destructive control for plurality and Condorcet voting and constructive and destructive control for approval voting. Plurality voting was the first natural system (among those having a polynomial-time winner problem) found to be fully resistant to candidate control. Faliszewski et al. [FHHR09a] studied the control complexity of the whole family of Llull/Copeland voting systems and were the first to find a natural voting system that is resistant to all common types of constructive control.[2] Another such voting system with full resistance to constructive control, and also to candidate control, is sincere-strategy preference-based approval voting (SP-AV), as shown by Erdély, Nowak, and Rothe [ENR09]. Prior to this paper, SP-AV was the voting system displaying the broadest control resistance among natural systems with a polynomial-time winner problem. However, SP-AV (as modified by Erdély, Nowak, and Rothe [ENR09]) is arguably less natural a system than fallback voting.[3] Note also that plurality has fewer resistances to voter control and Copeland voting has fewer resistances to destructive control than either of Bucklin and fallback voting. If we disregard SP-AV for the reasons mentioned in Footnote 3, all "natural" systems with polynomial-time winner determination (whose control behavior has been studied previously) are vulnerable to considerably more control types than Bucklin or fallback voting: plurality to six types, both Copeland and Condorcet to seven types, Llull to eight types, and approval voting to nine types. As mentioned above, shortly after our results were made public, Menton [Men12] showed that normalized range voting has the same number of resistances as fallback voting.

The parameterized complexity of electoral control has been studied by Betzler and Betzler and Uhlmann [BU09] and Faliszewski et al. [FHHR09a] for Llull/Copeland voting, by Liu et al. [LFZL09] for plurality, Condorcet, and approval voting, and by Liu and Zhu [LZ10] for maximin voting.

Faliszewski, Hemaspaandra, and Hemaspaandra [FHH11] study control for a more flexible type of attack (so-called "multimode control") and focus more on "combined" vulnerability than resistance. Meir et al. [MPRZ08] consider a different type of control scenario as well, using utility functions rather than constructive/destructive control, and they restrict their attention to adding/deleting candidates/voters. Both papers just mentioned study different voting systems. Recently, Hemaspaandra et al. [HHR12b, HHR12c] investigated *online* control in *sequential* elections, which is quite a different model than standard control in simultaneous elections studied here, see also their related work [HHR12a] on online manipulation in sequential elections.

As mentioned earlier, manipulation is related to, but different from control and has been studied even more extensively, in particular in [BTT89, BO91, CSL07, CS03, EL05, BEF11, OEH11,

---

[2]Hemaspaandra, Hemaspaandra, and Rothe [HHR09] construct, via "hybridization," a system with perfect control resistance. However, this system is artificial and shouldn't be used in practice, and it was not designed for that purpose.

[3]SP-AV is another hybrid system combining approval and preference-based voting; Brams and Sanver [BS06] proposed the original system and Erdély, Nowak, and Rothe [ENR09] its modification SP-AV. The reason why we said SP-AV is less natural than fallback voting is that, to preserve the votes' "admissibility" (as required by Brams and Sanver [BS06] to preclude trivial approval strategies), SP-AV employs an additional rule to (re-)coerce admissibility if in the course of a control action an originally admissible vote becomes inadmissible. As discussed in detail by Baumeister et al. [BEH+10], this rule, if applied, changes the approval strategies of the originally cast votes—a severe drawback. In contrast, here we study the *original fallback voting system* of Brams and Sanver [BS09] where votes, once cast, do not change.

FHS10, FHS08, IKM10, MR12, XZP$^+$09, XC08a, XC08b, FHHR11, BBHH10]. Elkind and Erdélyi [EE12] follow up the approach to consider voting rule uncertainty proposed by Baumeister, Roos, and Rothe [BRR11] for the more general possible winner problem. Much of this work has been surveyed in [FHH10, FP10, Con10, BEH$^+$10, FHHR09b]. Among the recent highlights regarding manipulation are the papers by Betzler, Niedermeier, and Woeginger [BNW11] and Davies et al. [DKNW11], who built on earlier work by Xia, Conitzer, and Procaccia [XCP10] to show that Borda voting is NP-hard to manipulate, even with only two manipulators and even in the unweighted case.

Note that all our NP-hardness results regard control problems with *unweighted votes* as well. It is worthwhile to explain the similarities and differences between manipulation and control at this example. Betzler, Niedermeier, and Woeginger [BNW11] describe the unweighted coalitional manipulation problem on p. 55 as follows: "Can one add a certain number of additional votes (called manipulators) to an election such that a distinguished candidate becomes a winner?" This problem is somewhat reminiscent of constructive control by adding voters. However, while the manipulators are free to choose their votes at will, the chair exerting control is restricted to a pool of given votes to choose from.[4]

**Organization:** This paper is organized as follows. In Section 2, we recall some notions from social choice theory, define the commonly studied types of control, and explain Bucklin voting and the fallback voting procedure of Brams and Sanver [BS09] in detail. Our results on the classical and parameterized control complexity of Bucklin voting and fallback voting are presented in Section 3. Finally, Section 4 provides some conclusions and open questions.

## 2 Preliminaries

### 2.1 Elections and Voting Systems

An *election* $(C,V)$ is given by a finite set $C$ of candidates and a finite list $V$ of votes over $C$. A *voting system* is a rule that specifies how to determine the winner(s) of any given election. The two voting systems considered in this paper are Bucklin voting and fallback voting. Bucklin voting is named after James W. Bucklin and was used from 1909 till 1922 in Grand Junction, Colorado [HH26]. Bucklin voting is therefore also referred to as Grand Junction voting. Between 1910 and 1917, it was also adopted in real elections in many other cities in the United States (see, e.g., http://www.electology.org/bucklin); although named "Bucklin voting," the system they used was actually what Brams and Sanver [BS09] introduced as fallback voting, a hybrid voting system that combines Bucklin with approval voting. For the social-choice properties of these systems, we refer to [HH26, BF78, BF83, BF02, RBLR11].

In *Bucklin voting*, votes are represented as (strict) linear orders over $C$, i.e., each voter ranks all candidates according to his or her preferences. For example, if $C = \{a,b,c,d\}$ then a vote might look like $c\ d\ a\ b$, i.e., this voter (strictly) prefers $c$ to $d$, $d$ to $a$, and $a$ to $b$. Given an election $(C,V)$ and a candidate $c \in C$, define the *level i score of c in* $(C,V)$ (denoted by $score^i_{(C,V)}(c)$) as the number of votes in $V$ that rank $c$ among their top $i$ positions. Denoting the *strict majority threshold for a list V of voters* by $maj(V) = \lfloor \|V\|/2 \rfloor + 1$, the *Bucklin score of c in* $(C,V)$ is the smallest $i$ such that $score^i_{(C,V)}(c) \geq maj(V)$. All candidates with a smallest Bucklin score, say $k$, and a largest level $k$ score are the *Bucklin winners (BV winners, for short) in* $(C,V)$. If some candidate becomes a Bucklin winner on level $k$, we call him or her a *level k BV winner in* $(C,V)$. Note that a level 1 BV winner must be unique, but there may be more level $k$ BV winners than one for $k > 1$, i.e., an election may have more than one Bucklin winner in general.

---

[4]Note also that for most voting systems the NP-hardness reductions for control by adding voters tend to be easier than those for control by partition of voters; compare, e.g., the proof of Theorem 3.14 with Theorems 3.22, 3.26, and 3.30 (the latter two using Constructions 3.24 and 3.28 and Lemmas 3.25 and 3.29).

In *approval voting*, votes are represented by approval vectors in $\{0,1\}^{\|C\|}$ (with respect to a fixed order of the candidates in $C$), where 0 stands for disapproval and 1 stands for approval. Given an election $(C,V)$ and a candidate $c \in C$, define the *approval score of c in* $(C,V)$ (denoted by $score_{(C,V)}(c)$) as the number of $c$'s approvals in $(C,V)$, and all candidates with a largest approval score are the *approval winners in* $(C,V)$. Note that an election may have more than one approval winner.

*Fallback voting* combines Bucklin with approval voting as follows. Each voter provides both an approval vector and a linear ordering of all approved candidates. The subset of candidates approved by a voter is also called his or her *approval strategy*. For simplicity, we will omit the disapproved candidates in each vote.[5] For example, if $C = \{a,b,c,d\}$ and a voter approves of $a$, $c$, and $d$ but disapproves of $b$, and prefers $c$ to $d$ and $d$ to $a$, then this vote will be written as: $c\ d\ a$. We will always explicitly state the candidate set, so it will always be clear which candidates participate in an election and which of them are disapproved by which voter (namely those not occurring in his or her vote). Given an election $(C,V)$ and a candidate $c \in C$, the notions of *level i score of c in* $(C,V)$ and *level k fallback voting winner (level k FV winner, for short) in* $(C,V)$ are defined analogously to the case of Bucklin voting, and if there exists a level $k$ FV winner for some $k \leq \|C\|$, he or she is called a *fallback winner (FV winner, for short) in* $(C,V)$. However, unlike in Bucklin voting, in fallback voting it may happen that no candidate reaches a strict majority for any level, due to voters being allowed to disapprove of (any number of) candidates, so it may happen that for no $k \leq \|C\|$ a level $k$ FV winner exists. In such a case, every candidate with a largest (approval) score is an *FV winner in* $(C,V)$. Note that Bucklin voting is the special case of fallback voting where each voter approves of all candidates.

As a notation, when a vote contains a subset of the candidate set, such as $c\ D\ a$ for a subset $D \subseteq C$, this is a shorthand for $c\ d_1\ \cdots\ d_\ell\ a$, where the elements of $D = \{d_1,\ldots,d_\ell\}$ are ranked with respect to some (tacitly assumed) fixed ordering of all candidates in $C$. For example, if $C = \{a,b,c,d\}$ is assumed to be ordered lexicographically and $D = \{b,d\}$ then "$c\ D\ a$" is a shorthand for the vote $c\ b\ d\ a$. If the candidate set is downsized in the process of a control action, the deleted candidates are removed in every vote, e.g., a vote $a\ c\ b\ d$ over the candidate set $C = \{a,b,c,d\}$ is altered to $c\ d$ if the subset $C' = \{c,d\}$ of candidates is considered.

## 2.2 Types of Electoral Control

There are eleven types of electoral control, each coming in two variants. In *constructive control* [BTT92], the chair tries to make his or her favorite candidate win; in *destructive control* [HHR07], the chair tries to prevent a despised candidate's victory. We refrain from giving a detailed discussion of natural, real-life scenarios for each of these 22 standard control types that motivate them; these can be found in, e.g., [BTT92, HHR07, FHHR09a, HHR09, ENR09]. However, we stress that every control type is motivated by an appropriate real-life scenario, and we will briefly point some of them out below.

---

[5]Erdélyi and Rothe [ER10] use a slightly different notation by separating the approved candidates from the disapproved candidates by a line, where all candidates to the left of this line are ranked and those to its right are unranked.

When we now formally define our 22 standard control types as decision problems, we assume that each election or subelection in these control problems will be conducted with the voting system at hand (i.e., either Bucklin or fallback voting) and that each vote will be represented as required by the corresponding voting system. We also assume that the chair has complete knowledge of the voters' preferences and/or approval strategies. This assumption may be considered to be unrealistic in certain settings, but is reasonable and natural in certain others, including small-scale elections among humans and even large-scale elections among software agents. More to the point, assuming the chair to have complete information makes sense for our results, as most of our results are NP-hardness lower bounds showing resistance of a voting system against specific control attempts and complexity lower bounds in the complete-information model are inherited by any natural partial-information model; see [HHR07] for a more detailed discussion of this point.

### 2.2.1 Control by Adding Candidates

We formally state our control problems in the common instance/question format. We start with the four problems modeling control by adding candidates. In these control scenarios, the chair seeks to make his or her favorite candidate win (in the constructive cases) or prevent a victory of his or her despised candidate (in the destructive cases) via introducing new candidates from a given pool of spoiler candidates into the election. Faliszewski et al. [FHHR09a] formalize this problem as follows. (In the definitions of control problems below, whenever we have a proper candidate subset $C' \subset C$ for an election $(C, V)$ then $(C', V)$ denotes the election where the voters in $V$ are restricted to $C'$.)

| CONSTRUCTIVE CONTROL BY ADDING A LIMITED NUMBER OF CANDIDATES (CCAC) |
|---|
| **Given:** An election $(C \cup D, V)$, $C \cap D = \emptyset$, a distinguished candidate $c \in C$, and a nonnegative integer $k$. ($C$ is the set of originally qualified candidates and $D$ is the set of spoiler candidates that may be added.) |
| **Question:** Does there exist a subset $D' \subseteq D$ such that $\|D'\| \leq k$ and $c$ is the unique winner (under the election system at hand) of election $(C \cup D', V)$? |

CONSTRUCTIVE CONTROL BY ADDING AN UNLIMITED NUMBER OF CANDIDATES, the problem variant originally proposed by Bartholdi et al. [BTT92], is the same except there is no limit $k$ on the number of spoiler candidates that may be added. We abbreviate this problem variant by CCAUC. Faliszewski et al. [FHHR09a] discuss in detail the reasons of why it makes sense to also consider the limited version of the problem. Although the difference in the definitions may appear to be negligible, note that the complexity of these problems differs significantly in some cases, e.g., in Llull's voting system [FHHR09a].

The destructive variants of both problems defined above are obtained by asking whether $c$ is *not* a unique winner of $(C \cup D', V)$. We use the shorthands DCAUC and DCAUC.

### 2.2.2 Control by Deleting Candidates

This control problem is defined analogously to control by adding a limited number of candidates, except that the chair now seeks to make a distinguished candidate $c$ win by deleting up to $k$ candidates from the given election.[6] This control scenario models candidate suppression. For example, by deleting certain candidates other than $c$ the chair may hope that their voters swing to now support $c$.

---

CONSTRUCTIVE CONTROL BY DELETING CANDIDATES (CCDC)

**Given:**   An election $(C,V)$, a distinguished candidate $c \in C$, and a nonnegative integer $k$.
**Question:**   Does there exist a subset $C' \subseteq C$ such that $\|C'\| \leq k$ and $c$ is the unique winner (under the election system at hand) of election $(C - C', V)$?

---

The destructive version of this problem is the same except that the chair now wants to preclude $c$ from being a unique winner (and, to prevent the problem from being trivial, simply deleting $c$ is not allowed). We use the shorthand DCDC.

### 2.2.3 Control by Partition or Run-Off Partition of Candidates

Both CONSTRUCTIVE CONTROL BY PARTITION OF CANDIDATES and CONSTRUCTIVE CONTROL BY RUN-OFF PARTITION OF CANDIDATES take as input an election $(C,V)$ and a candidate $c \in C$ and ask whether $c$ can be made the unique winner in a certain two-stage election consisting of one (in the partition case) or two (in the run-off partition case) first-round subelection(s) and a final round. In both variants, following Hemaspaandra, Hemaspaandra, and Rothe [HHR07], we consider two tie-handling rules, TP ("ties promote") and TE ("ties eliminate"), that enter into force when more candidates than one are tied for winner in any of the first-round subelections:

---

CONSTRUCTIVE CONTROL BY RUN-OFF PARTITION OF CANDIDATES WITH TP RULE (CCRPC-TP)

**Given:**   An election $(C,V)$ and a distinguished candidate $c \in C$.
**Question:**   Is it possible to partition $C$ into $C_1$ and $C_2$ such that $c$ is the unique winner (under the election system at hand) of election $(W_1 \cup W_2, V)$, where $W_i$, $i \in \{1,2\}$, is the set of winners of subelection $(C_i, V)$?

---

CONSTRUCTIVE CONTROL BY PARTITION OF CANDIDATES WITH TP RULE (CCPC-TP)

**Given:**   An election $(C,V)$ and a distinguished candidate $c \in C$.
**Question:**   Is it possible to partition $C$ into $C_1$ and $C_2$ such that $c$ is the unique winner (under the election system at hand) of election $(W_1 \cup C_2, V)$, where $W_1$ is the set of winners of subelection $(C_1, V)$?

---

In both cases, when the TE rule is used, none of multiple, tied first-round subelection winners is promoted to the final round. For example, if we have a run-off and $\|W_2\| \geq 2$ then the final-round election collapses to $(W_1, V)$; only a *unique* first-round subelection winner is promoted to the final round in TE. We abbreviate these two problem variants by CCRPC-TE and CCPC-TE. Note that the candidate set in the final round can be empty when the TE rule is used. In this case the resulting two-stage election has no winner. The following example gives a real-life scenario of a two-stage election as it can result from control by partition of candidates (although this is not an example of actually exerting this control type).

---

[6]No unlimited version has been considered previously for this control type or for the types of control by adding or deleting voters to be defined below.

**Example 2.1** *In the Eurovision Song Contest, which has been broadcast annually since 1956 on live television in Europe and other parts of the world (e.g., in more than 130 countries in 2006), each participating country submits a song (that has previously been selected in a national competition) and casts votes for the other countries' songs. The active member countries of the European Broadcasting Union (EBU) are eligible to participate in this competition. Since 2000, however, four EBU member countries have a privileged status because they are the four biggest financial contributors to the EBU: France, Germany, Spain, and the United Kingdom—the "Big Four"—are automatically qualified for the final round of the Eurovision Song Contest, whereas the other candidates have to participate in the semi-finals first to determine who among them enters the final round. Since 2011 Italy is also enjoying this privileged status, thus forming with the other countries the "Big Five." This is loosely reminiscent of a* CCPC-TP *scenario, where the participating countries are partitioned into* $C = C_1 \cup C_2$, $C_1$ *consisting of the semi-finalists and* $C_2$ *consisting of the Big Four/Five, such that all winners of the semi-finals (as modeled by the TP rule) move forward to the final round to run against the Big Four/Five.*

Other real-life examples include sports tournaments in which certain teams (such as last year's champion and the team hosting this year's championship) are given an exemption from qualification.

It is obvious how to obtain the destructive variants of these four problems formalizing control by candidate partition. We use the shorthands DCRPC-TP, DCPC-TP, DCRPC-TE, and DCPC-TE. Summing up, we now have defined 14 candidate control problems.

### 2.2.4 Control by Adding Voters

Turning now to the voter control problems, we start with control by adding voters. This control scenario models attempts by the chair to influence the outcome of elections via introducing new voters. There are many ways of introducing new voters into an election—think, for example, of "get-out-the-vote" drives, or of lowering the age-limit for the right to vote, or of attracting new voters with certain promises or even small gifts.

| CONSTRUCTIVE CONTROL BY ADDING VOTERS (CCAV) | |
|---|---|
| **Given:** | An election $(C, V \cup V')$, $V \cap V' = \emptyset$, where $V$ is a list of registered voters and $V'$ a pool of as yet unregistered voters that can be added, a distinguished candidate $c \in C$, and a nonnegative integer $k$. |
| **Question:** | Does there exist a sublist $V'' \subseteq V'$ of size at most $k$ such that $c$ is the unique winner (under the election system at hand) of election $(C, V \cup V'')$? |

The destructive variant of this problem is the same except that the chair now wants to preclude $c$ from being a unique winner. We use the shorthand DCAV.

### 2.2.5 Control by Deleting Voters

Disenfranchisement and other means of voter suppression is modeled as control by deleting voters.

| | |
|---|---|
| **CONSTRUCTIVE CONTROL BY DELETING VOTERS (CCDV)** | |
| **Given:** | An election $(C,V)$, a distinguished candidate $c \in C$, and a nonnegative integer $k$. |
| **Question:** | Does there exist a sublist $V' \subseteq V$ such that $\|V'\| \leq k$ and $c$ is the unique winner (under the election system at hand) of election $(C, V-V')$? |

Again, the destructive variant of this problem is the same except that the chair now wants to preclude $c$ from being a unique winner. We use the shorthand DCDV.

### 2.2.6 Control by Partition of Voters

| | |
|---|---|
| **CONSTRUCTIVE CONTROL BY PARTITION OF VOTERS (CCPV-TP)** | |
| **Given:** | An election $(C,V)$ and a distinguished candidate $c \in C$. |
| **Question:** | Is it possible to partition $V$ into $V_1$ and $V_2$ such that $c$ is the unique winner (under the election system at hand) of election $(W_1 \cup W_2, V)$, where $W_i$, $i \in \{1,2\}$, is the set of winners of subelection $(C, V_i)$? |

The destructive variant of this problem, denoted by DCPV-TP, is defined analogously, except it asks whether $c$ is *not* a unique winner of this two-stage election. In both variants, if one uses the tie-handling model TE instead of TP in the two first-stage subelections, a winner $w$ of $(C, V_1)$ or $(C, V_2)$ proceeds to the final stage if and only if $w$ is the only winner of his or her subelection. We use the shorthands CCPV-TE and DCPV-TE. Each of the four problems just defined models "two-district gerrymandering" (see e.g. [Bal08] for further information on district-gerrymandering in general).

Summing up, we now have defined eight voter control problems and thus a total of 22 control problems.

## 2.3 Classical and Parameterized Complexity

We assume the reader is familiar with the basic complexity classes such as P and NP. In classical complexity theory, a decision problem $A$ *(polynomial-time many-one) reduces* to a decision problem $B$ if there is a polynomial-time computable function $f$ such that for all inputs $x$, $x$ is a yes-instance for problem $A$ if and only if $f(x)$ is a yes-instance for problem $B$. A problem $B$ is NP-*hard* if every NP problem $A$ reduces to $B$, and $B$ is NP-*complete* if it is NP-hard and in NP. A problem $B$ is shown to be NP-hard by exhibiting a reduction to $B$ from a problem $A$ that is already known to be NP-hard. More background on (classical) complexity theory can be found, e.g., in the textbooks by Papadimitriou [Pap94] and Rothe [Rot05].

The theory of parameterized complexity was introduced by Downey and Fellows [DF99] based on a series of papers in the early 1990s and has developed into a vigorous branch of contemporary computer science, having strong applications in such areas as artificial intelligence and computational biology. The main idea is that for most NP-hard problems, typical inputs have secondary structure beyond the instance size measure $n$ that may significantly affect problem complexity in real-world computing contexts.

As a simple concrete illustration of the issue, the problem ML TYPE CHECKING, concerned with checking the consistency of type declarations in the ML programming language, was noted to be easy to solve in practice, despite being NP-hard. The explanation is that the relevant algorithm runs in time $O(2^k n)$, where $n$ is the instance size (here, the length of the ML program), and $k$ is the secondary measurement: the maximum nesting depth of the type declarations. For real-world ML programs, usually $k \leq 3$ and the algorithm easily solves the problem for typical instances.

This concrete example leads to the general setup of parameterized complexity theory. A *parameterized decision problem* is a decision problem in the classical sense, together with a specification of the secondary measurement (the *parameter*) of interest. The parameter may be an aggregate of several secondary measurements. The central notion is *fixed-parameter tractability* (FPT), meaning solvability in time $f(k)n^c$, where $f$ is an arbitrary function, and $c$ is a fixed constant. One can see that this generalizes polynomial time to this explicitly multivariate (two-dimensional) setting of parameterized decision problems.

There are some parameterized decision problems that seem not to admit fixed-parameter tractable algorithms. The well known graph problem VERTEX COVER, parameterized by the size of a solution set of vertices, is fixed-parameter tractable, with an algorithm having the same running time as in the above ML TYPE CHECKING example: $O(2^k n)$, where here $n$ is the number of vertices in the instance graph, and $k$ is the solution size. In contrast, the graph problem DOMINATING SET, parameterized by the size of a solution set of vertices, seems to admit no algorithm significantly better than $O(n^k)$, based on brute force examination of all up-to-$k$-subsets of the vertices.

Just as classical complexity is built from essentially four main ingredients:

- The central (desirable) notion of polynomial-time complexity.

- The notion of polynomial-time many-one reducibility that "transmits" the issue of polynomial-time solvability downward in the sense that if a decision problem $A$ reduces to a decision problem $B$, and if $B$ is polynomial-time solvable, then $A$ is as well.

- There is a hierarchy of classes of problems that are considered unlikely all to admit polynomial-time algorithms, e.g., NP and beyond; see, e.g., the textbook by Rothe [Rot05] for various hierarchies built upon NP.

- This consideration is supported by a highly plausible conjecture (namely, that P differs from NP) concerning the difficulty of analyzing the behavior of nondeterministic Turing machines.

Parameterized complexity is similarly structured:

- The central (desirable) outcome in the two-dimensional setting is fixed-parameter tractability.

- There is a corresponding notion of reduction between parameterized problems that transmits the issue of FPT solvability downward (see below for the definition).

- There is a hierarchy of presumably intractable parameterized problem classes; see below and, e.g., the monograph by Downey and Fellows [DF99].

- The presumption is underwritten by a highly plausible conjecture concerning the difficulty of analyzing the behavior of nondeterministic Turing machines, that is a natural (parameterized) variation on the central classical complexity conjecture.

**Definition 2.2 (Downey and Fellows [DF99])**    *1. A parameterized decision problem is a language $\mathscr{L} \subseteq \Sigma^* \times \mathbb{N}$. $\mathscr{L}$ is* fixed-parameter tractable *if there exists some computable function $f$ such that for each input $(x,k)$ of size $n = |(x,k)|$, it can be determined in time $\mathscr{O}(f(k) \cdot n^c)$ whether or not $(x,k)$ is in $\mathscr{L}$, where $c$ is a constant.*

2. *Given two parameterized problems $\mathscr{L}$ and $\mathscr{L}'$ (both encoded over $\Sigma^* \times \mathbb{N}$), we say $\mathscr{L}$ parameterized reduces to $\mathscr{L}'$ if there is a function $f : \Sigma^* \times \mathbb{N} \to \Sigma^* \times \mathbb{N}$ such that for each $(x,k)$,*

   (a) *$f(x,k) = (x',k')$ can be computed in time $\mathscr{O}(g(k) \cdot p(|x|))$ for some function $g$ and some polynomial $p$, and*

   (b) *$(x,k) \in \mathscr{L}$ if and only if $(x',k') \in \mathscr{L}'$, where $k' \leq g(k)$ (that is, $k'$ depends only on $k$).*

A parameterized problem $\mathscr{L}$ is *hard for a parameterized complexity class $\mathscr{C}$* if every problem in $\mathscr{C}$ parameterized reduces to $\mathscr{L}$, and $\mathscr{L}$ is *complete for $\mathscr{C}$* if it both belongs to $\mathscr{C}$ and is hard for $\mathscr{C}$.

The main hierarchy of parameterized complexity classes is:

$$\text{FPT} = \text{W}[0] \subseteq \text{W}[1] \subseteq \text{W}[2] \subseteq \cdots \subseteq \text{W}[t] \subseteq \cdots \subseteq \text{XP}.$$

FPT is the class of fixed-parameter tractable problems. W[1] is a strong parameterized analogue of NP, as the parameterized $k$-STEP HALTING PROBLEM FOR NONDETERMINISTIC TURING MACHINES is complete for W[1] under the above notion of parameterized reducibility [DF99]. Many classical decision problems have an obvious parameterized variant for a natural parameter. For example, the CLIQUE problem asks, given a graph $G$ and a positive integer $k$, whether $G$ has a clique (i.e., a subset of $G$'s vertices that are pairwise adjacent) of size at least $k$. In such cases, we let $k$-$\Pi$ denote this obvious parameterized variant of the classical decision problem $\Pi$. The parameterized $k$-CLIQUE problem is another problem complete for W[1], and the parameterized $k$-DOMINATING SET problem (which will be defined in Section 2.5) is complete for W[2]. The latter two parameterized problems (with parameter $k$, where $k$ is the solution size) are frequent sources of reductions that show likely parameterized intractability. XP is the class of parameterized decision problems solvable in time $O(n^{g(k)})$ for some function $g$.

See the monographs by Downey and Fellows [DF99], Niedermeier [Nie06], and Flum and Grohe [FG06] for further background on parameterized complexity theory and the rich toolkit of methods for devising fixes parameter tractable algorithms and proving parameterized hardness. Lindner and Rothe [LR08] survey a number of FPT and parameterized complexity results in computational social choice.

## 2.4   Immunity, Susceptibility, Resistance, and Vulnerability

Let $\mathfrak{CT}$ be a control type; for example, $\mathfrak{CT}$ might stand for "constructive control by partition of voters in model TP" or any of the other types of control defined in Section 2.2. We say a voting system is *immune to $\mathfrak{CT}$* if it is impossible for the chair to make the given candidate the unique winner in the constructive case (not a unique winner in the destructive case) via exerting control of type $\mathcal{CT}$. We say a voting system is *susceptible to* $\mathcal{CT}$ if it is not immune to $\mathcal{CT}$. A voting system that is susceptible to $\mathcal{CT}$ is said to be *vulnerable to* $\mathcal{CT}$ if the control problem corresponding to $\mathcal{CT}$ can be solved in polynomial time, and is said to be *resistant to* $\mathcal{CT}$ if the control problem corresponding to $\mathcal{CT}$ is NP-hard. These notions are due to Bartholdi et al. [BTT92] (except that we follow the now more common approach of Hemaspaandra, Hemaspaandra, and Rothe [HHR09] who define *resistant* to mean "susceptible and NP-hard" rather than "susceptible and NP-complete").

In analogy to the classical complexity notion, we say that a voting system is *parameterized-resistant to* $\mathcal{CT}$ *(with respect to a specified parameter)* if it is susceptible to $\mathcal{CT}$ and the parameterized decision problem corresponding to $\mathcal{CT}$ and this parameter is W[2]-hard. For example, a natural parameter to look at for control by deleting candidates is the number of candidates deleted.

## 2.5    Classical and Parameterized Decision Problems to Be Reduced From

In this section we introduce the decision problems that will be used for hardness proofs throughout this paper. We begin with the NP-complete problem EXACT COVER BY THREE-SETS, which will be used in Section 3.4 and is defined as follows (see, e.g., [GJ79]):

| | |
|---|---|
| | EXACT COVER BY THREE-SETS (X3C) |
| **Given:** | A set $B = \{b_1, b_2, \ldots, b_{3m}\}$, $m > 1$,[7] and a collection $\mathscr{S} = \{S_1, S_2, \ldots, S_n\}$ of subsets $S_i \subseteq B$ with $\|S_i\| = 3$ for each $i$, $1 \leq i \leq n$. |
| **Question:** | Is there a subcollection $\mathscr{S}' \subseteq \mathscr{S}$ such that each element of $B$ occurs in exactly one set in $\mathscr{S}'$? |

For the hardness proofs for several candidate control cases in Section 3.3 and one case of voter control in Section 3.4, we will use a restricted version of the NP-complete problem HITTING SET (see, e.g., [GJ79]), which is defined as follows:

| | |
|---|---|
| | RESTRICTED HITTING SET (RHS) |
| **Given:** | A set $B = \{b_1, b_2, \ldots, b_m\}$, a collection $\mathscr{S} = \{S_1, S_2, \ldots, S_n\}$ of nonempty subsets $S_i \subseteq B$ such that $n > m$, and a positive integer $k$ with $1 < k < m$. |
| **Question:** | Does $\mathscr{S}$ have a hitting set of size at most $k$, i.e., is there a set $B' \subseteq B$ with $\|B'\| \leq k$ such that for each $i$, $S_i \cap B' \neq \emptyset$? |

Note that by dropping the requirement "$n > m > k > 1$," we obtain the (unrestricted) HITTING SET problem. This restriction is needed to ensure that in the election constructed from an RHS instance in Construction 3.24 (see page 38), the scores of certain candidates are bounded. Construction 3.24 is an adaptation to Bucklin voting of a construction due to Hemaspaandra, Hemaspaandra, and Rothe [HHR07], namely Construction 4.28 in their paper, which they used to handle several candidate control cases for plurality voting. The election constructed there starts from a differently defined problem that is also called "RESTRICTED HITTING SET" but that is restrictive in another sense by requiring $n(k+1) \leq m - k$. In both constructions, the restriction serves the same purpose: bounding the scores of certain candidates so as to make the construction work.

Observe that our RESTRICTED HITTING SET problem is NP-complete as well, i.e., our restriction does not make the problem too easy. This is stated in the following lemma.

---

[7]Note that X3C is trivial to solve for $m = 1$.

**Lemma 2.3** RESTRICTED HITTING SET *is* NP-*complete.*

**Proof.** It is immediate that RESTRICTED HITTING SET is in NP. To show NP-hardness, we reduce the (general) HITTING SET problem to RESTRICTED HITTING SET. Let $(\hat{B}, \hat{\mathscr{S}}, \hat{k})$ be a given instance of HITTING SET, where $\hat{B} = \{b_1, b_2, \ldots, b_{\hat{m}}\}$ is a set, $\hat{\mathscr{S}} = \{S_1, S_2, \ldots, S_{\hat{n}}\}$ is a collection of nonempty subsets of $\hat{B}$, and $\hat{k} \leq \hat{m}$ is a positive integer. If $\hat{k} = \hat{m}$ or $\hat{k} = 1$, $(\hat{B}, \hat{\mathscr{S}}, \hat{k})$ is trivially in HITTING SET, so we may assume that $1 < \hat{k} < \hat{m}$.

Define the following instance $(B, \mathscr{S}, k)$ of RESTRICTED HITTING SET:

$$
\begin{aligned}
B &= \begin{cases} \hat{B} \cup \{a\} & \text{if } \hat{n} \leq \hat{m} \\ \hat{B} & \text{if } \hat{n} > \hat{m}, \end{cases} \\
\mathscr{S} &= \begin{cases} \hat{\mathscr{S}} \cup \{S_{\hat{n}+1}, S_{\hat{n}+2}, \ldots, S_{\hat{m}+2}\} & \text{if } \hat{n} \leq \hat{m} \\ \hat{\mathscr{S}} & \text{if } \hat{n} > \hat{m}, \end{cases} \\
k &= \begin{cases} \hat{k} + 1 & \text{if } \hat{n} \leq \hat{m} \\ \hat{k} & \text{if } \hat{n} > \hat{m}, \end{cases}
\end{aligned}
$$

where $S_{\hat{n}+1} = S_{\hat{n}+2} = \cdots = S_{\hat{m}+2} = \{a\}$.

Let $n$ be the number of members of $\mathscr{S}$ and $m$ be the number of elements of $B$. Since $1 < \hat{k} < \hat{m}$, we have $1 < k < m$. Note that if $\hat{n} > \hat{m}$ then $(B, \mathscr{S}, k) = (\hat{B}, \hat{\mathscr{S}}, \hat{k})$, so $n = \hat{n} > \hat{m} = m$; and if $\hat{n} \leq \hat{m}$ then $n = \hat{m} + 2 > \hat{m} + 1 = m$. Thus, in both cases $(B, \mathscr{S}, k)$ fulfills the restriction of RESTRICTED HITTING SET.

It is easy to see that $\hat{\mathscr{S}}$ has a hitting set of size at most $\hat{k}$ if and only if $\mathscr{S}$ has a hitting set of size at most $k$. In particular, assuming $\hat{n} \leq \hat{m}$, if $\hat{\mathscr{S}}$ has a hitting set $B'$ of size at most $\hat{k}$ then $B' \cup \{a\}$ is a hitting set of size at most $k = \hat{k} + 1$ for $\mathscr{S}$; and if $\hat{\mathscr{S}}$ has no hitting set of size at most $\hat{k}$ then $\mathscr{S}$ can have no hitting set of size at most $k = \hat{k} + 1$ (because $a \notin \hat{B}$, so $\{a\} \cap S_i = \emptyset$ for each $i$, $1 \leq i \leq \hat{n}$). $\qquad\square$

Regarding parameterized complexity, many W[2]-hardness results are proven via parameterized reductions from parameterized graph problems. We will prove W[2]-hardness of certain parameterized control problems via a parameterized reduction from the $k$-DOMINATING SET problem, which was shown to be W[2]-complete by Downey and Fellows [DF99]. Before we introduce this problem, we need some basic graph-theoretic notions.

**Definition 2.4** *Let $G = (B, A)$ be an undirected graph without loops or multiple edges.*[8]

*We say that two distinct vertices $b_i$ and $b_j$ are* adjacent *in $G$ if and only if there is an edge $\{b_i, b_j\} \in A$. Adjacent vertices are called* neighbors *in $G$.*

*The* neighborhood *of a vertex $b_i \in B$ is defined by $\mathrm{N}(b_i) = \{b_j \in B \mid \{b_i, b_j\} \in A\}$. The* closed neighborhood *of $b_i \in B$ is defined by $\mathrm{N}[b_i] = \mathrm{N}(b_i) \cup \{b_i\}$. For a subset $S \subseteq B$, the* neighborhood *of $S$ is defined as $\mathrm{N}(S) = \bigcup_{b_i \in S} \mathrm{N}(b_i)$ and the* closed neighborhood *of $S$ is defined as $\mathrm{N}[S] = \bigcup_{b_i \in S} \mathrm{N}[b_i]$.*

*A subset $B' \subseteq B$ is said to be a* dominating *set in $G$ if for each $b_i \in B - B'$ there is a $b_j \in B'$ such that $\{b_i, b_j\} \in A$. The* size *of a dominating set is the number of its vertices.*

---

[8]Note that we denote the vertex set of a graph not by $V$, as would be common, but rather by $B$, in order to avoid confusion with voter lists (for which "$V$" is reserved in this paper) and also in accordance with $B$ being the base set in the instances of the problems X3C and RHS.

Now we can define the W[2]-complete problem $k$-Dominating Set:

| $k$-Dominating Set ($k$-DS) | |
|---|---|
| **Given:** | A graph $G = (B, A)$ and a positive integer $k \le \|B\|$. |
| **Parameter:** | $k$. |
| **Question:** | Is there a dominating set of size at most $k$ in $G$? |

In other words, the (parameterized) dominating set problem tests, given a graph $G = (B, A)$ and an integer $k$ as the parameter, whether there is a subset $B' \subseteq B$ of size at most $k$ such that $B = \mathrm{N}[B']$. To distinguish the classical variant of this problem from its parameterized version just defined above, we drop the parameter in the problem name of the latter and simply write Dominating Set (DS), one of the standard NP-complete problems (see, e.g., [GJ79]).

Note that, without loss of generality, we can assume that $n > 2$ holds in any given DS and $k$-DS instance, respectively, since the thus restricted problem remains NP-complete and W[2]-complete, respectively.

**Remark 2.5** *If $\Pi$ is a (classical) decision problem in P, then $k$-$\Pi$ is in FPT for each parameter $k$. This gives a useful link between the two easiness notions in classical and parameterized complexity theory. In particular,* FPT *membership of a parameterized control problem (with respect to any parameter) follows from the voting system's vulnerability to the corresponding control type.*

*It may be tempting to assume there would be a similarly direct link between the hardness notions in classical and parameterized complexity theory, such as "W[2]-hardness of $k$-$\Pi$ immediately implies NP-hardness for $\Pi$." However, this statement is false, in general. The classical framework, which is built around P and reductions, and the parameterized framework, which is built around FPT and parameterized reductions, are essentially orthogonal. For concrete examples, with respect to suitable parameters,* VC-Dimension *(VCD) is unlikely to be NP-hard but $k$-VCD is complete for W[1], and* Tournament Dominating Set *(TDS) is unlikely to be NP-hard but $k$-TDS is W[2]-complete,[9] see [DF99] for more details. That being said, if one has a proof of, say, W[2]-hardness by a reduction from the NP- and W[2]-complete problems DS and $k$-DS, respectively, and the transformation is polynomial-time, then of course, the transformation shows both NP- and W[2]-hardness at the same time, and we will use this fact in some of our proofs.*

In the reductions to be presented in Sections 3.3 and 3.4, we will always start from a given instance of an NP-complete and/or W[2]-complete decision problem suitable for the control problem at hand, that is, we will always start from a given X3C, RHS, DS, or $k$-DS instance. In the upcoming constructions, the set of candidates in the elections to be defined will always contain the set $B$ from these instances. That is, for each element (or vertex) $b_i \in B$ there is a candidate $b_i$ in the election constructed, and it will always be clear from the context whether we mean an element (or a vertex) or a candidate when writing $b_i$. We will also refer to $S_i$ (or $\mathrm{N}[b_i]$) as a "subset of the candidates," namely, the set of candidates corresponding to the elements (or vertices) in $B$ that are in $S_i$ (or in vertex $b_i$'s closed neighborhood, $\mathrm{N}[b_i]$, $1 \le i \le n$). (Recall the sets $S_i$ from the definition of X3C and RHS and the notion of $\mathrm{N}[b_i]$ from Definition 2.4.)

---

[9]TDS is the problem of deciding whether a given tournament has a dominating set of size at most $k$ (the parameter in $k$-TDS). VCD is related to the so-called "Vapnik–Chervonenkis dimension" (which is the parameter in $k$-VCD), a problem quite central in learning theory; see [DF99] for the definition and more details.

| Control by | Fallback Voting | | Bucklin Voting | | SP-AV | | Approval Voting | |
|---|---|---|---|---|---|---|---|---|
| | Const. | Dest. | Const. | Dest. | Const. | Dest. | Const. | Dest. |
| Adding Candidates | **R** | **R** | **R** | **R** | R | R | I | V |
| Adding Candidates (unlimited) | **R***| **R***| **R***| **R***| R | R | I | V |
| Deleting Candidates | **R***| **R***| **R***| **R***| R | R | V | I |
| Partition of Candidates | **TE: R** | **TE: R** | **TE: R** | **TE: R** | TE: R | TE: R | TE:V | TE: I |
| | **TP: R** | **TP: R** | **TP: R** | **TP: R** | TP: R | TP: R | TP: I | TP: I |
| Run-off Partition of Candidates | **TE: R** | **TE: R** | **TE: R** | **TE: R** | TE: R | TE: R | TE:V | TE: I |
| | **TP: R** | **TP: R** | **TP: R** | **TP: R** | TP: R | TP: R | TP: I | TP: I |
| Adding Voters | **R***| **V** | **R***| **V** | R | V | R | V |
| Deleting Voters | **R***| **V** | **R***| **V** | R | V | R | V |
| Partition of Voters | **TE: R** | **TE: R***| **TE: R** | **TE: R***| TE: R | TE: V | TE: R | TE: V |
| | **TP: R** | **TP: R** | **TP: R** | **TP: S** | TP: R | TP: R | TP: R | TP: V |

Table 1: Overview of classical and parameterized complexity results for control in Bucklin and fallback voting. Key: I = immune, S = susceptible, R = resistant, R* = parameterized-resistant, V = vulnerable, TE = ties eliminate, and TP = ties promote. Results new to this paper are in boldface. Results for approval voting are due to [HHR07]. Results for SP-AV are due to [ENR09].

## 3 Control Complexity in Bucklin and Fallback Voting

### 3.1 Overview

Table 1 shows in boldface our results on the control complexity of Bucklin voting and fallback voting for all 22 standard control types. Since fallback voting combines Bucklin and approval voting, the table also shows the results for approval voting due to [HHR07]. [10] The other voting system displayed in the table, SP-AV, is yet another hybrid voting system combining approval with preference-based voting, which has been introduced by Brams and Sanver [BS06] and modified and studied in [ENR09]. Until now, with 19 out of 22 resistances SP-AV has been the system with the most known resistances to electoral control among natural voting systems with polynomial-time winner determination. Recall from Remark 2.5 that for every R* entry in the table (i.e., for every W[2]-hardness with respect to the parameters stated later in our theorems and corollaries), we will show a corresponding resistance (i.e., NP-hardness) result by essentially the same reduction. Thus, fallback voting has 20 and Bucklin voting has 19 out of 22 resistances, so fallback voting has one more resistance to electoral control than SP-AV and Bucklin voting at least draws level. More importantly, however, as argued in the introduction, both fallback voting and Bucklin voting are much more natural voting systems than SP-AV as modified by Erdélyi, Nowak, and Rothe [ENR09].

Table 2 gives an overview of the reductions used to prove the new (i.e., boldfaced) resistance and parameterized resistance results in Table 1. The first column states the problem from which we reduce. The second column states the control problem shown to be (parameterized) resistant, where the problem name has the prefix "BV" if this control problem refers to Bucklin voting (which immediately implies hardness also for fallback voting). The only problem name with the prefix "FV" for fallback voting, FV-DCPV-TP in the last row, refers to the case where the complexity is open for Bucklin voting. The third column of Table 2 points at the corresponding theorem or construction and the fourth column states whether it is a parameterized reduction or not.

---

[10]Note that the results for control by adding a limited number of candidates in approval voting, though not explicitly considered in [HHR07], can be obtained straightforwardly from their proofs for the corresponding "unlimited" adding-candidates case.

| Reduction from | to | Reference | parameterized? |
|---|---|---|---|
| $k$-DS | BV-CCDC | Theorem 3.5 | yes |
| | BV-DCDC | Theorem 3.6 | |
| | BV-CCAV | Theorem 3.14 | |
| | BV-CCDV | Theorem 3.18 | |
| | BV-CCAC | Theorem 3.8 | |
| | BV-DCAC | | |
| | BV-CCAUC | | no |
| | BV-DCAUC | | |
| | BV-DCPV-TE | Construction 3.24 and Theorem 3.26 | yes |
| RHS | BV-CCPC-TE | Construction 3.10 and Theorem 3.12 | no |
| | BV-CCPC-TP | | |
| | BV-DCPC-TE | | |
| | BV-DCPC-TP | | |
| | BV-CCRPC-TE | | |
| | BV-CCRPC-TP | | |
| | BV-DCRPC-TE | | |
| | BV-DCRPC-TP | | |
| | FV-DCPV-TP | Construction 3.28 and Theorem 3.30 | |
| X3C | BV-CCPV-TE | Theorem 3.22 | |
| | BV-CCPV-TP | | |

Table 2: Overview of the reductions used to prove the new results in Table 1.

## 3.2 Susceptibility

If an election system $\mathscr{E}$ satisfies the "unique" variant of the Weak Axiom of Revealed Preference[11] (Unique-WARP, for short), then $\mathscr{E}$ is immune to constructive control by adding candidates (no matter whether a limited or an unlimited number of candidates is being added), and this observation has been applied to approval voting [BTT92, HHR07]. Unlike approval voting, however, Bucklin voting and fallback voting do not satisfy Unique-WARP.

**Proposition 3.1** *Neither Bucklin voting nor fallback voting satisfies Unique-WARP.*

**Proof.** We show this result for Bucklin voting only; the proof for fallback voting follows immediately. Consider the election $(C, V)$ with candidate set $C = \{a, b, c, d\}$ and voter collection

---

[11]This variant of the axiom says that the unique winner $w$ of any election is also the unique winner of every subelection including $w$.

$V = (v_1, v_2, \ldots, v_6)$:

| | $(C,V)$ | | | | | | $a$ | $b$ | $c$ | $d$ |
|---|---|---|---|---|---|---|---|---|---|---|
| $v_1 = v_2 = v_3$ : | $a$ | $c$ | $b$ | $d$ | $score^1_{(C,V)}$ : | | 3 | 2 | 0 | 1 |
| $v_4 = v_5$ : | $b$ | $d$ | $c$ | $a$ | $score^2_{(C,V)}$ : | | **4** | 2 | 3 | 3 |
| $v_6$ : | $d$ | $a$ | $c$ | $b$ | | | | | | |

Candidate $a$ is the unique Bucklin winner of election $(C,V)$, reaching the strict majority threshold on level 2 with $score^2_{(C,V)}(a) = 4$. By removing candidate $b$ from the election, we get the subelection $(C',V)$ with $C' = \{a,c,d\}$.

| | $(C',V)$ | | | | | | $a$ | $c$ | $d$ |
|---|---|---|---|---|---|---|---|---|---|
| $v_1 = v_2 = v_3$ : | $a$ | $c$ | $d$ | | $score^1_{(C',V)}$ : | | 3 | 0 | 3 |
| $v_4 = v_5$ : | $d$ | $c$ | $a$ | | $score^2_{(C',V)}$ : | | 4 | **5** | 3 |
| $v_6$ : | $d$ | $a$ | $c$ | | | | | | |

There is no candidate on level 1 who passes the strict majority threshold. However, there are two candidates on the second level with a strict majority, namely candidates $a$ and $c$. Since $score^2_{(C',V)}(c) = 5 > 4 = score^2_{(C',V)}(a)$, the unique Bucklin winner of subelection $(C',V)$ is candidate $c$. Thus, Bucklin voting does not satisfy Unique-WARP. ❑

Indeed, as we will now show, Bucklin voting and fallback voting are susceptible to each of our 22 control types. Our proofs make use of the results of [HHR07] that provide general proofs of and links between certain susceptibility cases, which we will here refer to as Theorems HHR07-4.1, HHR07-4.2, and HHR07-4.3. For the sake of self-containment, Figure 1, which is taken from [RBLR11, p. 199], gives an overview of the susceptibility links in various control scenarios from these three theorems of Hemaspaandra et al. [HHR07]. An arrow between two control types in this figure, say $\mathfrak{CT}_1 \to \mathfrak{CT}_2$ means that susceptibility to $\mathfrak{CT}_1$ implies susceptibility to $\mathfrak{CT}_2$; and the arrows between control types and the two properties in dashed boxes mean that (1) susceptibility to destructive control by deleting candidates implies that Unique-WARP is violated, and (2) every voiced[12] election system is susceptible to destructive control by adding candidates. We start with susceptibility to candidate control for Bucklin voting.

**Lemma 3.2** *Bucklin voting is susceptible to constructive and destructive control by adding candidates (in both the "limited" and the "unlimited" case), by deleting candidates, and by partition of candidates (with or without run-off and for each in both model TE and model TP).*

**Proof.** From parts 1 and 2 of Theorem HHR07-4.3 (see also Figure 1) and the fact that Bucklin voting is a voiced voting system, it follows that Bucklin voting is susceptible to constructive control by deleting candidates and to destructive control by adding candidates (in both the "limited" and the "unlimited" case).

Now, consider the election $(C,V)$ given in the proof of Proposition 3.1. The unique Bucklin winner of the election is candidate $a$. Partition $C$ into $C_1 = \{a,c,d\}$ and $C_2 = \{b\}$. The unique

---

[12]An election system is said to be *voiced* if the single candidate in any one-candidate election always wins.

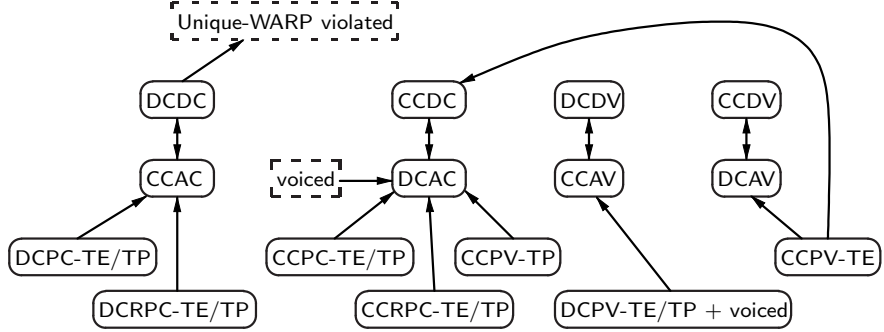

Figure 1: Susceptibility links in various control scenarios due to [HHR07]

Bucklin winner of subelection $(C_1, V)$ is candidate $c$, as shown in the proof of Proposition 3.1. In both partition and run-off partition of candidates and for each in both tie-handling models, TE and TP, candidate $b$ runs against candidate $c$ in the final stage of the election.

$$
\begin{array}{ll}
& \underline{(\{b,c\}, V)} \qquad\qquad\qquad \underline{b \quad c} \\
v_1 = v_2 = v_3 : & c \ b \qquad\qquad score^1_{(\{b,c\}, V)} : \quad 2 \ \ \mathbf{4} \\
v_4 = v_5 : & b \ c \\
v_6 : & c \ b
\end{array}
$$

The unique Bucklin winner is in each case candidate $c$. Thus, Bucklin voting is susceptible to destructive control by partition of candidates (with or without run-off and for each in both model TE and model TP).

By part 4 of Theorem HHR07-4.2 (see also Figure 1), Bucklin voting is also susceptible to destructive control by deleting candidates. By part 1 of Theorem HHR07-4.1 (see also Figure 1), Bucklin voting is also susceptible to constructive control by adding candidates (in both the "limited" and the "unlimited" case).

Changing the roles of $a$ and $c$ makes $c$ our distinguished candidate. In election $(C, V)$, $c$ loses against candidate $a$. By partitioning the candidates as described above, $c$ becomes the unique Bucklin winner of the election. Thus, Bucklin voting is susceptible to constructive control by partition of candidates (with or without run-off and for each in both tie-handling models, TE and TP). ❑

We now turn to susceptibility to voter control for Bucklin voting.

**Lemma 3.3** *Bucklin voting is susceptible to constructive and destructive control by adding voters, by deleting voters, and by partition of voters (in both model TE and model TP).*

**Proof.** Consider election $(C,V)$, where $C = \{a,b,c,d\}$ is the set of candidates and $V = (v_1,v_2,v_3,v_4)$ is the collection of voters with the following preferences:

| $(C,V)$ | | | | | | $a$ | $b$ | $c$ | $d$ |
|---|---|---|---|---|---|---|---|---|---|
| $v_1:$ | $a$ | $c$ | $b$ | $d$ | $score^1_{(C,V)}:$ | 1 | 2 | 0 | 1 |
| $v_2:$ | $d$ | $c$ | $a$ | $b$ | $score^2_{(C,V)}:$ | **3** | 2 | 2 | 1 |
| $v_3:$ | $b$ | $a$ | $c$ | $d$ | | | | | |
| $v_4:$ | $b$ | $a$ | $c$ | $d$ | | | | | |

Clearly, candidate $a$ is the unique Bucklin winner of $(C,V)$ on the second level. We partition $V$ into $V_1 = (v_1,v_2)$ and $V_2 = (v_3,v_4)$. Thus we split $(C,V)$ into two subelections:

| $(C,V_1)$ | | | | | and | $(C,V_2)$ | | | | | | $a$ | $b$ | $c$ | $d$ | | | $a$ | $b$ | $c$ | $d$ |
|---|---|---|---|---|---|---|---|---|---|---|---|---|---|---|---|---|---|---|---|---|---|
| $v_1:$ | $a$ | $c$ | $b$ | $d$ | | | | | | | $score^1_{(C,V_1)}:$ | 1 | 0 | 0 | 1 | $score^1_{(C,V_2)}:$ | 0 | **2** | 0 | 0 |
| $v_2:$ | $d$ | $c$ | $a$ | $b$ | | | | | | | $score^2_{(C,V_1)}:$ | 1 | 0 | **2** | 1 | | | | | |
| $v_3:$ | | | | | | | $b$ | $a$ | $c$ | $d$ | | | | | | | | | | |
| $v_4:$ | | | | | | | $b$ | $a$ | $c$ | $d$ | | | | | | | | | | |

However, $c$ is the unique Bucklin winner of $(C,V_1)$ and $b$ is the unique Bucklin winner of $(C,V_2)$, and so $a$ is not promoted to the final stage. Thus, Bucklin voting is susceptible to destructive control by partition of voters in both tie-handling models, TE and TP.

With this and by part 1 of Theorem HHR07-4.3 (see also Figure 1) and the fact that Bucklin voting is a voiced voting system, Bucklin voting is susceptible to destructive control by deleting voters. By part 3 of Theorem HHR07-4.1 (see also Figure 1), Bucklin voting is also susceptible to constructive control by adding voters.

By changing the roles of $a$ and $c$ again, we can see that Bucklin voting is susceptible to constructive control by partition of voters in both model TE and model TP. By part 3 of Theorem HHR07-4.2 (see also Figure 1), Bucklin voting is also susceptible to constructive control by deleting voters. Finally, again by part 4 of Theorem HHR07-4.1 (see also Figure 1), Bucklin voting is susceptible to destructive control by adding voters. ❑

Since Bucklin voting is a special case of fallback voting, fallback voting is also susceptible to all 22 common types of control.

**Corollary 3.4** *Fallback voting is susceptible to each of the 22 control types defined in Section 2.2.*

### 3.3 Candidate Control

In this section, we show the (parameterized) resistance results for candidate control in Bucklin and fallback voting. Recall from Section 2.4 that "parameterized resistance" (indicated by the R* entries in Table 1) refers to W[2]-hardness of the corresponding control problem with respect to a specified parameter, and "resistance" (indicated by the R entries in Table 1) refers to NP-hardness of the corresponding control problem. Also, recall from Section 2.5 that in the reductions to be presented in this section, the elections constructed will always contain a subset $B$ of candidates, where each candidate $b_i$ corresponds to the element $b_i$ from the set $B$ given in the (parameterized) decision problem the reduction starts from. It will always be clear from the context whether a candidate or an element is meant by $b_i$.

### 3.3.1 Control by Deleting Candidates

**Theorem 3.5** *Bucklin voting is resistant to constructive control by deleting candidates, and is parameterized-resistant when this control problem is parameterized by the number of candidates deleted.*

**Proof.** In light of Remark 2.5 it is enough to give just one reduction to prove both claims at the same time. Let $(G,k)$ with $G = (B,E)$ be a given instance of $k$-DOMINATING SET as described in Section 2.5. Without loss of generality, we may assume that $k < n = \|B\|$, since the set $B$ of all vertices trivially is a dominating set in $G$.

Define the election $(C,V)$, where $C = B \cup D \cup \{w\} \cup X \cup Y$ is the set of candidates, $w$ is the distinguished candidate, $D$ is a set of "co-winners" (see below), and $X$ and $Y$ are sets of *padding candidates*.[13]

**Co-winners in $D$:** $D$ is a set of $k+1$ candidates that tie with $w$. These candidates prevent that deleting up to $k$ co-winners of $(C,V)$ makes $w$ the unique winner.

**Padding candidates in $X$:** $X$ is a set of $n(n+k) - \sum_{i=1}^{n} \|N[b_i]\|$ candidates such that for each $i$, $1 \le i \le n$, we can find a subset $X_i \subseteq X$ with $n+k-\|N[b_i]\|$ elements such that $X_i \cap X_j = \emptyset$ for all $i,j \in \{1,\dots,n\}$ with $i \ne j$. These subsets ensure that $w$ is always placed at the $(n+k+1)$st position in the first voter group of $V$ below.

**Padding candidates in $Y$:** $Y$ is a set of $n(k+1)$ candidates such that for each $j$, $1 \le i \le k+1$, we can find a subset $Y_j \subseteq Y$ with $n$ elements such that $Y_i \cap Y_j = \emptyset$ for all $i,j \in \{1,\dots,k+1\}$ with $i \ne j$. These subsets ensure that each $d_j \in D$ is always placed at the $(n+k+1)$st position in the second voter group of $V$ below.

$V$ is the following collection of $2n$ voters, so that we have a strict majority with $n+1$ votes:

| #  | For each ...           | number of votes | ranking of candidates                                            |
|----|------------------------|-----------------|------------------------------------------------------------------|
| 1  | $i \in \{1,\dots,n\}$  | 1               | $N[b_i]$ $X_i$ $w$ $((B-N[b_i]) \cup (X-X_i) \cup Y)$ $D$         |
| 2  | $j \in \{1,\dots,k+1\}$| 1               | $Y_j$ $(D-\{d_j\})$ $d_j$ $(B \cup X \cup (Y-Y_j) \cup \{w\})$    |
| 3  |                        | $n-k-1$         | $D$ $(X \cup Y \cup \{w\})$ $B$                                   |
| 4  |                        | 1               | $D$ $w$ $(B \cup X \cup Y)$                                       |
| 5  |                        | 1               | $X$ $Y$ $B$ $(D \cup \{w\})$                                      |

Note that when up to $k$ candidates are deleted (no matter which ones), the candidates from $D$ can never be among the top $n+k$ candidates in the votes of the first voter group. Table 3 shows the scores on the relevant levels of the relevant candidates in election $(C,V)$.

Note that the candidates in $D$ and candidate $w$ are the only level $n+k+1$ Bucklin winners of election $(C,V)$, since there is no other candidate reaching a strict majority of $n+1$ votes or more on any level up to $n+k+1$.

---

[13]Note that in this construction as well as in later constructions, the subsets of padding candidates are always constructed so as to ensure that, at least up to a certain level, no padding candidate scores enough points to be relevant for the outcome of the election. So in the following argument the padding candidates are mainly ignored and their scores are not listed in the overview tables.

|  | $b_i \in B$ | $w$ | $d_j \in D$ |
|---|---|---|---|
| $score^{k+1}$ | $\leq n$ | $0$ | $n-k$ |
| $score^{k+2}$ | $\leq n$ | $1$ | $n-k$ |
| $score^{n+k}$ | $\leq n$ | $1$ | $n$ |
| $score^{n+k+1}$ | $\leq n$ | $\mathbf{n+1}$ | $\mathbf{n+1}$ |

Table 3: Level $i$ scores in $(C,V)$ for $i \in \{k+1, k+2, n+k, n+k+1\}$ and the candidates in $C - (X \cup Y)$.

We claim that $G$ has a dominating set of size $k$ if and only if $w$ can be made the unique Bucklin winner by deleting at most $k$ candidates.

From left to right: Suppose $G$ has a dominating set $B' \subseteq B$ of size $k$. Delete the corresponding candidates from $C$. Since $B'$ is a dominating set in $G$ (i.e., $B = \mathrm{N}[B']$), every $b_i \in B$ has a neighbor in $B'$ or is itself in $B'$, which means that in election $(C - B', V)$ candidate $w$ gets pushed at least one position to the left in each of the $n$ votes in the first voter group. So $w$ reaches a strict majority already on level $n+k$ with a score of $n+1$. Since no other candidate does so (in particular, no candidate in $D$), it follows that $w$ is the unique level $n+k$ Bucklin winner of $(C - B', V)$.

From right to left: Suppose $w$ can be made the unique Bucklin winner of the election by deleting at most $k$ candidates. Since there are $k+1$ candidates other than $w$ (namely, those in $D$) having a strict majority on level $n+k+1$ in election $(C, V)$, deleting $k$ candidates from $D$ is not sufficient for making $w$ the unique Bucklin winner of the resulting election. So by deleting at most $k$ candidates, $w$ must become the unique Bucklin winner on a level lower than or equal to $n+k$. This is possible only if $w$ is pushed at least one position to the left in all votes from the first voter group. This, however, implies that the $k' \leq k$ deleted candidates are either

1. all contained in $B$ and correspond to a dominating set of size $k'$ for $G$, or

2. are in $B \cup X$.

Note that not all deleted candidates can be contained in $X$, since $k < n$ and the sets $X_i$, $1 \leq i \leq n$, are pairwise disjoint. If some of the $k'$ deleted candidates are in $X$, say $\ell < k'$ of them, let $B'$ be the set containing the $k' - \ell$ other candidates that have been deleted. For each $i$, $1 \leq i \leq n$, if in the $i$th voter of the first group no candidate from $\mathrm{N}[b_i]$ was deleted but a candidate $x_j$ from $X_i$, add an arbitrary candidate from $\mathrm{N}[b_i]$ to $B'$ instead of $x_j$. This yields again a dominating set of size $k'$ for $G$. In both cases, if $k' < k$ then by adding $k - k'$ further candidates from $B$ (which is possible due to $k < n$) we obtain a dominating set of size $k$ for $G$.

Note that this polynomial-time reduction is parameterized, as the given parameter $k$ of $k$-Dominating Set is the same parameter $k$ that bounds the number of candidates allowed to be deleted in the control problem. ❏

**Theorem 3.6** *Bucklin voting is resistant to destructive control by deleting candidates, and is parameterized-resistant when this control problem is parameterized by the number of candidates deleted.*

**Proof.** In light of Remark 2.5 it is again enough to give just one reduction to prove both claims at the same time. For the W[2]-hardness proof in the destructive case for Bucklin voting, let $(G, k)$ with $G = (B, E)$ be a given instance of $k$-Dominating Set. Define the election $(C, V)$, where

$$C = \{c, w\} \cup B \cup M_1 \cup M_2 \cup M_3 \cup X \cup Y \cup Z$$

is the candidate set, $c$ is the distinguished candidate, and $M_1, M_2, M_3, X, Y$, and $Z$ are sets of padding candidates (recall Footnote 13).

**Padding candidates in $M_1, M_2$, and $M_3$:** $M_1$, $M_2$, and $M_3$ are three pairwise disjoint sets, where each is a set of $k$ candidates that are positioned in the votes so as to ensure that no other candidate besides $w$ and $c$ can reach a strict majority up to level $n + k$.

**Padding candidates in $X$:** $X$ is a set of $n^2 - \sum_{i=1}^{n} \|\mathrm{N}[b_i]\|$ candidates such that for each $i$, $1 \leq i \leq n$, we can find a subset $X_i \subseteq X$ with $n - \|\mathrm{N}[b_i]\|$ elements such that $X_i \cap X_j = \emptyset$ for all $i, j \in \{1, \ldots, n\}$ with $i \neq j$. These subsets ensure that $w$ is always placed at the $(n+1)$st position in the first voter group of $V$ below.

**Padding candidates in $Y$:** $Y$ is a set of $n - 1$ padding candidates ensuring that $c$ is at position $n$ in the votes of the second voter group of $V$ below.

**Padding candidates in $Z$:** $Z$ is a set of $n - 2$ padding candidates ensuring that $w$ is at position $n - 1$ and $c$ is at position $n$ in the vote of the third voter group of $V$ below.

$V$ is the following collection of $2n + 1$ voters, so we have a strict majority threshold of $n + 1$:

| # | For each … | number of votes | ranking of candidates |
|---|---|---|---|
| 1 | $i \in \{1, \ldots, n\}$ | 1 | $\mathrm{N}[b_i] \, X_i \, w \, M_1 \, ((B - \mathrm{N}[b_i]) \cup M_2 \cup M_3 \cup (X - X_i) \cup Y \cup Z) \, c$ |
| 2 | | $n$ | $Y \, c \, M_2 \, (B \cup M_1 \cup M_3 \cup X \cup Z \cup \{w\})$ |
| 3 | | 1 | $Z \, w \, c \, M_3 \, (B \cup M_1 \cup M_2 \cup X \cup Y)$ |

| | $b_i \in B$ | $w$ | $c$ |
|---|---|---|---|
| $score^{n-1}$ | $\leq n$ | 1 | 0 |
| $score^{n}$ | $\leq n$ | 1 | $\mathbf{n+1}$ |
| $score^{n+1}$ | $\leq n$ | $n+1$ | $n+1$ |

Table 4: Level $i$ scores in $(C, V)$ for $i \in \{n-1, n, n+1\}$ and the candidates in $B \cup \{c, w\}$.

Table 4 gives an overview of the scores on the relevant levels of the relevant candidates in election $(C, V)$. Note that candidate $c$ is the unique level $n$ Bucklin winner of election $(C, V)$, since $c$ is the first candidate reaching a strict majority of votes (namely, $n + 1$ points on level $n$, as indicated—here and in later score tables as well—by a boldfaced entry).

We claim that $G$ has a dominating set of size $k$ if and only if $c$ can be prevented from being a unique Bucklin winner by deleting at most $k$ candidates.

From left to right: Suppose $G$ has a dominating set $B' \subseteq B$ of size $k$. Delete the corresponding candidates. Now candidate $w$ moves at least one position to the left in each of the $n$ votes in the first voter group. Since candidate $c$ reaches a strict majority no earlier than on level $n$ and $score^n_{(C-B',V)}(w) = n+1 = score^n_{(C-B',V)}(c)$, candidate $c$ is no longer a unique Bucklin winner of the resulting election.

From right to left: Suppose $c$ can be prevented from being a unique Bucklin winner of the election by deleting at most $k$ candidates. Note that deleting one candidate from an election can move the strict majority level of another candidate at most one level to the left. Observe that only candidate $w$ can prevent $c$ from winning the election, since $w$ is the only candidate other than $c$ who reaches a strict majority of votes until level $n+k$. In election $(C, V)$, candidate $w$ reaches this majority no earlier than on level $n+1$, and candidate $c$ not before level $n$. Thus $w$ can prevent $c$ from being a unique winner only by scoring at least as many points as $c$ no later than on level $n$. This is possible only if $w$ is pushed at least one position to the left in all votes of the first voter group. By an argument analogous to that given in the constructive case for this control type (see the proof of Theorem 3.5), this implies that $G$ has a dominating set of size $k$.

Note that this polynomial-time reduction is parameterized, as the given parameter $k$ of $k$-DOMINATING SET is the same parameter $k$ that bounds the number of candidates allowed to be deleted in the control problem. ❏

**Corollary 3.7** *Fallback voting is resistant to constructive and destructive control by deleting candidates, and is parameterized-resistant when these two control problems are parameterized by the number of candidates deleted.*

### 3.3.2 Control by Adding Candidates

**Theorem 3.8** *Bucklin voting is resistant to constructive and destructive control by adding an unlimited and a limited number of candidates, and is parameterized-resistant when the limited variants are parameterized by the number of candidates added.*

**Proof.** We begin with the limited cases and note that again, in light of Remark 2.5 it suffices to prove W[2]-hardness; NP-hardness follows by the same reduction. We do so first for constructive control by adding a limited number of candidates. Let $(G, k)$ with $G = (B, E)$ be a given instance of $k$-DOMINATING SET. Recall from Section 2.5 that without loss of generality we can assume that $n > 2$ holds. Define the election $(C, V)$, where $C = \{c, w\} \cup X \cup Y \cup Z$ with $X = \{x_1, x_2, \ldots, x_{n-1}\}$, $Y = \{y_1, y_2, \ldots, y_{n-2}\}$, and $Z = \{z_1, z_2, \ldots, z_{n-1}\}$ is the set of candidates, $B$ is the set of spoiler candidates, $w$ is the distinguished candidate, and $V$ is the following collection of $2n+1$ voters:

| # | For each … | number of votes | ranking of candidates |
|---|---|---|---|
| 1 | $i \in \{1, \ldots, n\}$ | 1 | $N[b_i] \ X \ c \ ((B - N[b_i]) \cup Y \cup Z \cup \{w\})$ |
| 2 | | $n$ | $Y \ c \ w \ (B \cup X \cup Z)$ |
| 3 | | 1 | $Z \ w \ c \ (B \cup X \cup Y)$ |

Note that the candidate subsets $X$, $Y$, and $Z$ each contain padding candidates ensuring that candidates $w$ and $c$ do not reach a strict majority of votes on a level lower than $n$. Table 5 shows the relevant scores on all relevant levels in both elections, $(C, V)$ and $(C \cup B, V)$, but not the scores of the padding candidates (recall Footnote 13). Note that there is no boldfaced entry for election $(C \cup B, V)$ in this table, as we here give only an overview of the relevant scores; whether or not $w$ is a unique winner on level $n$ depends on the exact score of $c$, which in turn depends on the structure of the graph $G$.

|  | $(C,V)$ | | $(C \cup B, V)$ | | |
|---|---|---|---|---|---|
|  | $w$ | $c$ | $b_i \in B$ | $w$ | $c$ |
| $score^{n-1}$ | $0$ | $n$ | $\leq n$ | $0$ | $n$ |
| $score^n$ | $n+1$ | $\mathbf{2n}$ | $\leq n$ | $n+1$ | $n$ |
| $score^{n+1}$ | $n+1$ | $2n+1$ | $\leq 2n+1$ | $n+1$ | $\leq 2n+1$ |

Table 5: Level $i$ scores in $(C,V)$ and $(C \cup B, V)$ for $i \in \{n-1, n, n+1\}$ and all relevant candidates.

The strict majority threshold is reached with $n+1$ points, regardless of the candidate set considered (i.e., for both $(C,V)$ and $(C \cup B, V)$). Note that candidate $w$ is not a (unique) Bucklin winner of election $(C,V)$, since only candidates $c$ and $w$ can reach a strict majority up to level $n$ (actually, they do so *exactly* on level $n$), and it holds that

$$score^n_{(C,V)}(w) = n+1 < 2n = score^n_{(C,V)}(c).$$

Since $n > 2$, $c$ is the unique level $n$ Bucklin winner of $(C,V)$.

We claim that $G$ has a dominating set of size $k$ if and only if $w$ can be made the unique Bucklin winner by adding at most $k$ spoiler candidates from $B$ to $C$.

From left to right: Suppose $G$ has a dominating set $B'$ of size $k$. Add the corresponding candidates to $C$. Since $B'$ is a dominating set in $G$ (i.e., $B = N[B']$), every $b_i \in B$ has a neighbor in $B'$ or is itself in $B'$, which means that in election $(C \cup B', V)$ candidate $c$ gets pushed at least one position to the right in each of the $n$ votes of the first voter group. Thus, $c$ loses $n$ points on level $n$, which implies $score^n_{(C \cup B', V)}(c) = n$. It follows that $w$ is the unique Bucklin winner of $(C \cup B', V)$, since $w$ is the only candidate on level $n$ who has a strict majority of $n+1$ out of $2n+1$ votes in this election.

From right to left: Suppose $w$ can be made the unique Bucklin winner by adding some subset $B' \subseteq B$ of spoiler candidates with $\|B'\| \leq k \leq \|B\|$ to $C$. By adding any candidates from $B$ to the election, only votes in the first voter group can be affected in the first $n$ levels. Note that candidate $c$ has already on level $n-1$ a score of $n$ from the second voter group. Thus, to make $w$ (which has no more than $n+1$ approvals up to level $n$, whatever subset of $B$ is added) the unique Bucklin winner, $c$ must not gain any more points until level $n$ (otherwise, $c$ would tie with or beat $w$ on level $n$ with $score^n_{(C \cup B', V)}(c) \geq n+1$). This, however, is possible only if candidate $c$ is pushed in *all* votes from the first voter group at least one position to the right. This, in turn, is possible only if $B'$ is a dominating set in $G$. Since $k \leq \|B\|$, we can add any $k - \|B'\|$ elements from $B - B'$ to $B'$ if $\|B'\| < k$, yielding again a dominating set of $G$, so $G$ has a dominating set of size $k$.

For the W[2]-hardness proof in the destructive case, we have to do only one minor change: We switch the roles of candidates $c$ and $w$, i.e., now $c$ (instead of $w$) is the distinguished candidate and the chair wishes to dethrone $c$. Other than that, the instance of the destructive control problem to be constructed from a given instance of $k$-DOMINATING SET will be defined exactly as in the constructive case above. Now, by an analogous argument as in the constructive case, one can show that $G$ has a dominating set of size $k$ if and only if $c$ can be prevented from being a unique Bucklin winner by adding at most $k$ candidates.

Note that both reductions are parameterized, as the given parameter $k$ of $k$-DOMINATING SET is the same parameter $k$ that bounds the number of candidates allowed to be added in the two control problems.

The unlimited cases can be shown by the same reductions but for these cases the reductions are not parameterized, so for these two problems we have NP-hardness. ❑

Again, since Bucklin voting is a special case of fallback voting, we have the following corollary.

**Corollary 3.9** *Fallback voting is resistant to constructive and destructive control by adding an unlimited and a limited number of candidates, and is parameterized-resistant when the limited variants are parameterized by the number of candidates added.*

### 3.3.3 Control by Partition of Candidates

Construction 3.10 will be applied to prove the remaining eight cases of candidate control, establishing resistance (not parameterized resistance) in each case. This construction adapts Construction 4.28 of [HHR07], which they used to handle certain candidate control cases for plurality voting.[14] Note that they start from a HITTING SET instance, while we reduce from a restricted version of this problem, and in our construction there is one additional candidate, $d$, and there are two more voter groups than in their construction. These modifications are needed because Bucklin voting is more involved than plurality voting.

**Construction 3.10** *Let* $(B, \mathscr{S}, k)$ *be a given instance of* RESTRICTED HITTING SET*, where* $B = \{b_1, b_2, \ldots, b_m\}$ *is a set,* $\mathscr{S} = \{S_1, S_2, \ldots, S_n\}$ *is a collection of nonempty subsets* $S_i \subseteq B$ *such that* $n > m$, *and* $k < m$ *is a positive integer. (Thus,* $n > m > k > 1$.)

*Define the election* $(C, V)$, *where* $C = B \cup \{c, d, w\}$ *is the candidate set and where* $V$ *consists of the following* $6n(k+1) + 4m + 11$ *voters:*

| # | For each … | number of voters | ranking of candidates |
|---|---|---|---|
| 1 | | $2m+1$ | $c\ d\ B\ w$ |
| 2 | | $2n + 2k(n-1) + 3$ | $c\ w\ d\ B$ |
| 3 | | $2n(k+1)+5$ | $w\ c\ d\ B$ |
| 4 | $i \in \{1, \ldots, n\}$ | $2(k+1)$ | $d\ S_i\ c\ w\ (B - S_i)$ |
| 5 | $j \in \{1, \ldots, m\}$ | $2$ | $d\ b_j\ w\ c\ (B - \{b_j\})$ |
| 6 | | $2(k+1)$ | $d\ w\ c\ B$ |

The proof of Theorem 3.12, which establishes the remaining eight cases of candidate control for Bucklin voting, will make use of Lemma 3.11 below, which is based on Claims 4.29 and 4.30 of [HHR07] but is tailored to Bucklin instead of plurality voting. As Bucklin voting is more involved than plurality, so are our arguments in the proof of Lemma 3.11.

**Lemma 3.11** *Consider the election* $(C, V)$ *constructed according to Construction 3.10 from a* RESTRICTED HITTING SET *instance* $(B, \mathscr{S}, k)$.

---

[14]Their construction was also useful in the proofs of most candidate control results for SP-AV [ENR09], so the structure of the constructions and the arguments in the proofs of Lemma 3.11 and Theorem 3.12 are adaptations of those by [HHR07] and [ENR09], tailored here to Bucklin voting.

1. $c$ is the unique level 2 BV winner of $(\{c,d,w\},V)$.

2. If $\mathscr{S}$ has a hitting set $B'$ of size $k$, then $w$ is the unique BV winner of election $(B' \cup \{c,d,w\},V)$.

3. Let $D \subseteq B \cup \{d,w\}$. If $c$ is not a unique BV winner of election $(D \cup \{c\},V)$, then there exists a set $B' \subseteq B$ such that

   (a) $D = B' \cup \{d,w\}$,
   (b) $w$ is a level 2 BV winner of election $(B' \cup \{c,d,w\},V)$, and
   (c) $B'$ is a hitting set for $\mathscr{S}$ of size at most $k$.

**Proof.** For the first part, note that there is no level 1 BV winner in election $(\{c,d,w\},V)$ and we have the following level 2 scores in this election:

$$
\begin{aligned}
score^2_{(\{c,d,w\},V)}(c) &= 6n(k+1) + 2(m-k) + 9, \\
score^2_{(\{c,d,w\},V)}(d) &= 2n(k+1) + 4m + 2k + 3, \\
score^2_{(\{c,d,w\},V)}(w) &= 4n(k+1) + 2m + 10.
\end{aligned}
$$

Since $n > m$ (which implies $n > k$), we have:

$$
\begin{aligned}
score^2_{(\{c,d,w\},V)}(c) - score^2_{(\{c,d,w\},V)}(d) &= 4n(k+1) - (2m+4k) + 6 > 0, \\
score^2_{(\{c,d,w\},V)}(c) - score^2_{(\{c,d,w\},V)}(w) &= 2n(k+1) - (2k+1) > 0.
\end{aligned}
$$

Thus, $c$ is the unique level 2 BV winner of $(\{c,d,w\},V)$.

For the second part, suppose that $B'$ is a hitting set for $\mathscr{S}$ of size $k$. Then there is no level 1 BV winner in election $(B' \cup \{c,d,w\},V)$, and we have the following level 2 scores:

$$
\begin{aligned}
score^2_{(B' \cup \{c,d,w\},V)}(c) &= 4n(k+1) + 2(m-k) + 9, \\
score^2_{(B' \cup \{c,d,w\},V)}(d) &= 2n(k+1) + 4m + 2k + 3, \\
score^2_{(B' \cup \{c,d,w\},V)}(w) &= 4n(k+1) + 2(m-k) + 10, \\
score^2_{(B' \cup \{c,d,w\},V)}(b_j) &\leq 2n(k+1) + 2 \quad \text{for all } b_j \in B'.
\end{aligned}
$$

It follows that $w$ is the unique level 2 BV winner of election $(B' \cup \{c,d,w\},V)$.

For the third part, let $D \subseteq B \cup \{d,w\}$. Suppose $c$ is not a unique BV winner of election $(D \cup \{c\},V)$.

(3a) Besides $c$, only $w$ has a strict majority of votes on the second level and only $w$ can tie or beat $c$ in $(D \cup \{c\},V)$. Thus, since $c$ is not a unique BV winner of election $(D \cup \{c\},V)$, $w$ is clearly in $D$. In $(D \cup \{c\},V)$, candidate $w$ has no level 1 strict majority and candidate $c$ has already on level 2 a strict majority. Thus, $w$ must tie or beat $c$ on level 2. For a contradiction, suppose $d \notin D$. Then

$$
\begin{aligned}
score^2_{(D \cup \{c\},V)}(c) &\geq 4n(k+1) + 2m + 11; \\
score^2_{(D \cup \{c\},V)}(w) &= 4n(k+1) + 2m + 10,
\end{aligned}
$$

which contradicts the above observation that $w$ ties with or beats $c$ on level 2. Thus, $D = B' \cup \{d,w\}$, where $B' \subseteq B$.

(3b) This part follows immediately from the proof of part (3a).

(3c) Let $\ell$ be the number of sets in $\mathscr{S}$ not hit by $B'$. We have that

$$
\begin{aligned}
score^2_{(B' \cup \{c,d,w\}, V)}(w) &= 4n(k+1) + 10 + 2(m - \|B'\|), \\
score^2_{(B' \cup \{c,d,w\}, V)}(c) &= 2(m - k) + 4n(k+1) + 9 + 2(k+1)\ell.
\end{aligned}
$$

From part (3b) we know that

$$
score^2_{(B' \cup \{c,d,w\}, V)}(w) \geq score^2_{(B' \cup \{c,d,w\}, V)}(c),
$$

so

$$
4n(k+1) + 10 + 2(m - \|B'\|) \geq 2(m - k) + 4n(k+1) + 9 + 2(k+1)\ell.
$$

The above inequality implies

$$
1 > \frac{1}{2} \geq \|B'\| - k + (k+1)\ell.
$$

Since $T = \|B'\| - k + (k+1)\ell$ is an integer, we have $T \leq 0$. If $T = 0$ then $\ell = 0$ and $\|B'\| = k$. Now assume $T < 0$. If $\ell = 0$, $B'$ is a hitting set with $\|B'\| < k$, and if $\ell > 0$ then $(k+1)\ell > k$, which contradicts $T = \|B'\| - k + (k+1)\ell < 0$. In each possible case, we have a hitting set (as $\ell = 0$) of size at most $k$. This completes the proof of Lemma 3.11.                    ❏

Now we are ready to handle the eight remaining cases of candidate control.

**Theorem 3.12** *Bucklin voting is resistant to constructive and destructive control by partition of candidates and by run-off partition of candidates (for each in both tie-handling models, TE and TP).*

**Proof.**    Susceptibility holds by Lemma 3.2, so it remains to show NP-hardness. For the constructive cases, map the given RESTRICTED HITTING SET instance $(B, \mathscr{S}, k)$ to the election $(C, V)$ from Construction 3.10 with $w$ being the distinguished candidate.

We claim that $\mathscr{S}$ has a hitting set of size at most $k$ if and only if $w$ can be made the unique BV winner by exerting control via any of our four control scenarios (partition of candidates with or without run-off, and for each in either tie-handling model, TE and TP).

From left to right: Suppose $\mathscr{S}$ has a hitting set $B' \subseteq B$ of size $k$. Partition the set of candidates into the two subsets $C_1 = B' \cup \{c, d, w\}$ and $C_2 = C - C_1$. According to Lemma 3.11, $w$ is the unique level 2 BV winner of subelection $(C_1, V) = (B' \cup \{c, d, w\}, V)$. No matter whether we have a run-off or not, and regardless of the tie-handling rule used, the opponents of $w$ in the final stage (if there are any opponents at all) each are candidates from $B$. Since $n > m$, $w$ has a majority in the final stage on the first level with a score of $4n(k+1) + 9$. Thus, $w$ is the unique BV winner of the resulting election.

From right to left: Suppose $w$ can be made the unique BV winner via any of our four control scenarios. Since $c$ is not a BV winner of the election, there is a subset $D \subseteq B \cup \{d, w\}$ of candidates such that $c$ is not a unique BV winner of election $(D \cup \{c\}, V)$. By Lemma 3.11, there exists a hitting set for $\mathscr{S}$ of size at most $k$.

For the four destructive cases, we simply change the roles of $c$ and $w$ in the above argument. ❏

Again, since Bucklin voting is a special case of fallback voting, we have the following corollary.

**Corollary 3.13** *Fallback voting is resistant to constructive and destructive control by partition of candidates and by run-off partition of candidates (for each in both tie-handling models, TE and TP).*

## 3.4 Voter Control

Turning now to voter control for Bucklin voting and fallback voting, we start with control by adding and deleting voters, where we have both (classical) resistance and parameterized resistance results. As in the previous section, recall that "parameterized resistance" (the R* entries in Table 1) refers to W[2]-hardness of the corresponding control problem with respect to a specified parameter, and that "resistance" (the R entries in Table 1) refers to NP-hardness of the corresponding control problem. Again, recall from Section 2.5 that in the reductions to be presented in this section, the elections constructed will always contain a subset $B$ of candidates, where each candidate $b_i$ corresponds to the element $b_i$ from the set $B$ given in the (parameterized) decision problem the reduction starts from, and it will always be clear from the context whether $b_i$ is meant to be a candidate or an element.

### 3.4.1 Control by Adding Voters

**Theorem 3.14** *Bucklin voting is resistant to constructive control by adding voters, and is parameterized-resistant when this control problem is parameterized by the number of voters added.*

**Proof.** Again, in light of Remark 2.5 it is enough to give one reduction that proves both claims at the same time. Let $(G,k)$ with $G = (B,E)$ be a given instance of $k$-DOMINATING SET as described above. Define the election $(C,V \cup U)$, where $C = \{c,w\} \cup B \cup X \cup Y$ is the set of candidates, $w$ is the distinguished candidate, and $X$ and $Y$ are sets of padding candidates (recall Footnote 13 on page 21).

**Padding candidates in $X$:** $X$ is a set of $\sum_{i=1}^{n} \|N[b_i]\|$ candidates such that for each $i$, $1 \le i \le n$, we can find a subset $X_i \subseteq X$ with $\|N[b_i]\|$ elements such that $X_i \cap X_j = \emptyset$ for all $i,j \in \{1,\dots,n\}$ with $i \ne j$. These subsets ensure that $w$ is always placed at the $(n+1)$st position in the votes of the unregistered voters in $U$ below.

**Padding candidates in $Y$:** $Y$ is a set of $n$ padding candidates ensuring that none of the candidates in $B$ is ranked among the first $n$ candidates in the votes of the registered voters in $V$ below.

$V$ is the collection of registered and $U$ is the collection of unregistered voters. $V \cup U$ consists of the following $n+k-1$ voters:

| Voter list | For each … | number of voters | ranking of candidates |
|---|---|---|---|
| $V$ | | $k-1$ | $c$ $Y$ $B$ $w$ $X$ |
| $U$ | $i \in \{1,\dots,n\}$ | 1 | $(B - N[b_i])$ $X_i$ $w$ $c$ $(N[b_i] \cup (X - X_i) \cup Y)$ |

Clearly, $c$ is the unique level 1 Bucklin winner of election $(C,V)$.

We claim that $G$ has a dominating set of size $k$ if and only if $w$ can be made the unique Bucklin winner by adding at most $k$ voters from $U$.

From left to right: Suppose $G$ has a dominating set $B'$ of size $k$. Add the corresponding voters from $U$ to the election (i.e., each voter $u_i$ for which $b_i \in B'$). Now we have an election with $2k-1$ voters, so the strict majority threshold is $k$.

Since $B'$ is a dominating set, we have $B = N[B']$, so for each $b_j \in B$ there is at least one of the added voters $u_i$ such that $b_j \in N[b_i]$, which means that $b_j$ is ranked to the right of $w$ in these $k$ added votes. It follows that up to level $n+1$ only candidate $w$ will reach this threshold of $k$, hence $w$ is the unique Bucklin winner of this election.

From right to left: Suppose $w$ can be made the unique Bucklin winner by adding at most $k$ voters from $U$. Denote the set of these voters by $U'$ and note that $\|U'\| \leq k$. Note further that $score^1_{(C, V \cup U')}(c) = score^1_{(C, V)}(c) = k - 1$, that is, $c$ reaches a score of $k - 1$ already on the first level (with or without adding $U'$). However, if any candidate has a strict majority already on the first level, then he or she is the unique Bucklin winner of the election. As $w$ is the unique Bucklin winner of $(C, V \cup U')$, the strict majority threshold for $V \cup U'$ must be greater than $k - 1$. This, in turn, implies $\|U'\| \geq k$, so $\|U'\| = k$ and the strict majority threshold for $V \cup U'$ is exactly $k$. Note that $score^{n+1}_{(C, V \cup U')}(w) = k > k - 1 = score^{n+1}_{(C, V \cup U')}(x)$ and $score^n_{(C, V \cup U')}(w) = 0$. Moreover, since adding the voters from $U'$ to the election has made $w$ the unique Bucklin winner of $(C, V \cup U')$, none of the candidates in $B$ can be ranked among the first $n$ candidates by each voter in $U'$; otherwise (i.e., if some candidate $b_j \in B$ would be ranked among the first $n$ candidates by each voter in $U'$), we would have $score^n_{(C, V \cup U')}(b_j) = k$, i.e., $b_j$ would reach a strict majority in $(C, V \cup U')$ earlier than $w$, a contradiction. But this means that the voters in $U'$ correspond to a dominating set of size $k$ in $G$.

Note that this polynomial-time reduction is parameterized, as the given parameter $k$ of $k$-DOMINATING SET is the same parameter $k$ that bounds the number of voters allowed to be added in this control problem. ❑

**Corollary 3.15** *Fallback voting is resistant to constructive control by adding voters, and is parameterized-resistant when this control problem is parameterized by the number of voters added.*

In contrast to Corollary 3.15, fallback voting is vulnerable to destructive control by adding voters and, as we will see later in Theorem 3.20, it is also vulnerable to destructive control by deleting voters. In fact, the proof of Theorem 3.16 shows something slightly stronger: Fallback voting is what Hemaspaandra, Hemaspaandra, and Rothe [HHR07] call "certifiably vulnerable" to this type of destructive voter control, i.e., the algorithm showing vulnerability to destructive control by adding voters not only decides whether or not control is possible but it even computes a successful control action if one exists. Note that in the proof of Theorem 3.16, after giving a high-level description of the algorithm, we present the algorithm in detail and argue for its correctness in parallel. Therefore, we refrain from giving a formal presentation (e.g., in pseudocode, which would be overly difficult to read). Note further that the algorithm is not designed so as to optimizing its runtime (as then it might be harder to see it is correct); rather, we focus on clarity regarding the arguments for correctness. Thus, we give an upper bound for the runtime (with respect to the number of candidates in the given election) that still might be improved.

**Theorem 3.16** *Fallback voting is vulnerable to destructive control by adding voters, via an algorithm whose worst-case runtime is in $\mathscr{O}(m^2(n' + p(n+n')))$, where $m$ is the number of candidates, $n$, respectively $n'$, is the number of votes in $V$, respectively $V'$, and $p(x)$ is a polynomial giving an upper bound for the time needed to compute level scores in an election with $x$ voters.*

**Proof.** Susceptibility holds by Lemma 3.3. We present a polynomial-time algorithm for solving the destructive control by adding voters case. We will make use of the following notation. Given an election $(C,V)$, recall that $maj(V) = \lfloor \|V\|/2 \rfloor + 1$ denotes the strict majority threshold for $V$, and define the deficit of candidate $d \in C$ for reaching a strict majority in $(C,V)$ on level $i$, $1 \le i \le \|C\|$, by

$$def^i_{(C,V)}(d) = maj(V) - score^i_{(C,V)}(d).$$

The input to our algorithm is an election $(C, V \cup V')$ (where $C$ is the set of candidates, $V$ is the collection of registered voters, and $V'$ is the collection of unregistered voters), a distinguished candidate $c \in C$, and an integer $\ell$ (the number of voters allowed to be added). The algorithm either outputs a sublist $V'' \subseteq V'$, $\|V''\| \le \ell$, that describes a successful control action (if any exists), or indicates that control is impossible for this input.

We give a high-level description of the algorithm. We assume that $c$ is initially the unique FV winner of election $(C,V)$; otherwise, the algorithm simply outputs $V'' = \emptyset$ and halts, since there is no need to add any voters from $V'$.

Let $k$ be the largest number of candidates any voter in $V \cup V'$ approves of. Clearly, $k \le \|C\|$. The algorithm proceeds in at most $k + 1$ stages, where the last stage is the *approval stage* and checks whether $c$ can be dethroned as a unique FV winner by approval score via adding at most $\ell$ voters from $V'$, and all preceding stages are *majority stages* that check whether a candidate $d \ne c$ can tie or beat $c$ on level $i$ via adding at most $\ell$ voters from $V'$. Since the first majority stage is slightly different from the subsequent majority stages, we describe both cases separately.

**Majority Stage 1:** For each candidate $d \in C - \{c\}$, check whether $d$ can tie or beat $c$ on the first level via adding at most $\ell$ voters from $V'$. To this end, find a list $V'_d \subseteq V'$ of largest cardinality such that $\|V'_d\| \le \ell$ and all voters in $V'_d$ approve of $d$ on the first level. Check whether

$$(3.1) \qquad score^1_{(C,V \cup V'_d)}(d) \ \ge \ score^1_{(C,V \cup V'_d)}(c).$$

If (3.1) fails to hold, this $d$ is hopeless, so go to the next candidate (or to the next stage if all other candidates have already been checked in this stage). If (3.1) holds, check whether $d$ has a strict majority in $(C, V \cup V'_d)$ on the first level, and if so, output $V'' = V'_d$ and halt. Otherwise, this $d$ is hopeless, so go to the next candidate (or stage).

**Majority Stage $i$, $1 < i \le k$:** This stage is entered only if it was not possible to find a successful control action in majority stages $1, \ldots, i-1$. For each candidate $d \in C - \{c\}$, check whether $d$ can tie with or beat $c$ up to the $i$th level via adding at most $\ell$ voters from $V'$. To this end, find a list $V'_d \subseteq V'$ of largest cardinality such that $\|V'_d\| \le \ell$ and all voters in $V'_d$ approve of $d$ up to the $i$th level but disapprove of $c$ up to the $i$th level. Check whether

$$(3.2) \qquad score^i_{(C,V \cup V'_d)}(d) \ \ge \ score^i_{(C,V \cup V'_d)}(c).$$

If (3.2) fails to hold, this $d$ is hopeless, so go to the next candidate (or to the next stage if all other candidates have already been checked in this stage). If (3.2) holds, check whether $d$ has a strict majority in $(C, V \cup V'_d)$ on the $i$th level, and if so, check whether

$$(3.3) \qquad score^{i-1}_{(C,V \cup V'_d)}(c) \ \ge \ maj(V \cup V'_d).$$

If (3.3) fails to hold, output $V'' = V'_d$ and halt. Otherwise (i.e., if (3.3) holds), though $d$ might dethrone $c$ by adding $V'_d$ on the $i$th level, it was not quick enough: $c$ has already won earlier. In that case, find a largest list $V'_{cd} \subseteq V'$ such that

1. $\|V'_d \cup V'_{cd}\| \leq \ell$,

2. all voters in $V'_{cd}$ approve of both $c$ and $d$ up to the $i$th level, and

3. the voters in $V'_{cd}$ are chosen such that $c$ is approved of as late as possible by them (i.e., at levels with a largest possible number, where ties may be broken arbitrarily).

Now, check whether

$$(3.4) \qquad score^{i-1}_{(C, V \cup V'_d \cup V'_{cd})}(c) \quad \geq \quad maj(V \cup V'_d \cup V'_{cd}).$$

If (3.4) holds, then $d$ is hopeless, so go to the next candidate (or to the next stage if all other candidates have already been checked in this stage). Otherwise (i.e., if (3.4) fails to hold), check whether $\|V'_{cd}\| \geq def^i_{(C, V \cup V'_d)}(d)$. If so (note that $d$ has now a strict majority on level $i$), output $V'' = V'_d \cup V'_{cd}$ and halt. Note that, by choice of $V'_{cd}$, (3.2) implies that

$$score^i_{(C, V \cup V'_d \cup V'_{cd})}(d) \geq score^i_{(C, V \cup V'_d \cup V'_{cd})}(c).$$

Thus, in $(C, V \cup V'_d \cup V'_{cd})$, $d$ ties with or beats $c$ and has a strict majority on the $i$th level (and now, we are sure that $d$ was not too late). Otherwise (i.e., if $\|V'_{cd}\| < def^i_{(C, V \cup V'_d)}(d))$, this $d$ is hopeless, so go to the next candidate (or stage).

**Approval Stage:** This stage is entered only if it was not possible to find a successful control action in majority stages $1, 2, \ldots, k$.

First, check if

$$(3.5) \qquad score_{(C, V)}(c) < \left\lfloor \frac{\|V\| + \ell}{2} \right\rfloor + 1.$$

If (3.5) fails to hold, output "control impossible" and halt, since we have found no candidate in the majority stages who could tie or beat $c$ and would have a strict majority when adding at most $\ell$ voters from $V'$, so adding any choice of at most $\ell$ voters from $V'$ would $c$ still leave a strict majority. If (3.5) holds, looping over all candidates $d \in C - \{c\}$, check whether there are $score_{(C, V)}(c) - score_{(C, V)}(d) \leq \ell$ voters in $V'$ who approve of $d$ and disapprove of $c$. If this is not the case, move on to the next candidate, since $d$ could never catch up on $c$ via adding at most $\ell$ voters from $V'$. If it is the case for some $d \in C - \{c\}$, however, add this list of voters (call it $V'_d$) and check whether

$$(3.6) \qquad score_{(C, V \cup V'_d)}(c) \quad < \quad maj(V \cup V'_d).$$

If (3.6) holds, output $V'' = V'_d$ and halt. Otherwise (i.e., if (3.6) fails to hold), check whether

$$(3.7) \qquad \begin{aligned} \ell - \|V'_d\| &\geq \|V'_\emptyset\| \\ &\geq 2\left(score_{(C, V \cup V'_d)}(c) - \frac{\|V \cup V'_d\|}{2}\right), \end{aligned}$$

where $V'_\emptyset$ consists of those voters in $V'$ who disapprove of both candidates, $c$ and $d$. If (3.7) does not hold, move on to the next candidate, since after adding these voters $c$ would still have a strict majority. Otherwise (i.e., if (3.7) holds), add exactly $2(score_{(C,V \cup V'_d)}(c) - \|V \cup V'_d\|/2)$ voters from $V'_\emptyset$ (denoted by $V'_{\emptyset,+}$). Output $V'' = V'_d \cup V'_{\emptyset,+}$ and halt.

If we have entered the approval stage (because we were not successful in any of the majority stages), but couldn't find any candidate here who was able to dethrone $c$ by adding at most $\ell$ voters from $V'$, we output "control impossible" and halt.

The correctness of the algorithm follows from the remarks made above. Crucially, note that the algorithm proceeds in the "safest way possible": If there is any successful control action then our algorithm finds some successful control action.

It is also easy to see that this algorithm runs in polynomial time. More specifically, the algorithm proceeds in at most $k+1 \leq m+1$ stages, where in each stage certain conditions are checked for up to $m-1$ candidates. For checking the conditions we have to go through the list $V'$ and compute level scores over at most $n+n'$ voters, each a constant number of times. Thus, the algorithm's worst-case runtime is in $\mathcal{O}(m^2(n' + p(n+n')))$. (Recall that we didn't optimize it in terms of its runtime; rather, we described it in a way to make it easier to check its correctness.) ❑

Since Bucklin voting is a special case of fallback voting, Bucklin voting inherits vulnerability from fallback voting in this control scenario.

**Corollary 3.17** *Bucklin voting is vulnerable to destructive control by adding voters.*

### 3.4.2 Control by Deleting Voters

**Theorem 3.18** *Bucklin voting is resistant to constructive control by deleting voters, and is parameterized-resistant when this control problem is parameterized by the number of voters deleted.*

**Proof.** In light of Remark 2.5, we again give only one reduction that proves both claims at the same time. To prove W[2]-hardness, we again provide a reduction from $k$-DOMINATING SET. Let $(G,k)$ with $G = (B,E)$ be a given instance of this problem as described at the start of this section. Define the election $(C,V)$, where $C = \{c,w\} \cup B \cup X \cup Y \cup Z$ is the set of candidates, $w$ is the distinguished candidate, and $X$, $Y$, and $Z$ are sets of padding candidates (recall Footnote 13).

**Padding candidates in $X$:** $X$ is a set of $\sum_{i=1}^{n} \|B - N[b_i]\|$ candidates such that for each $i$, $1 \leq i \leq n$, we can find a subset $X_i \subseteq X$ with $n - \|N[b_i]\|$ elements such that $X_i \cap X_j = \emptyset$ for all $i,j \in \{1,\ldots,n\}$ with $i \neq j$. These subsets ensure that $c$ is always placed among the top $(n+1)$ positions in the first voter group of $V$ below.

**Padding candidates in $Y$:** $Y$ is a set of $\sum_{i=1}^{n} \|N[b_i]\|$ candidates such that for each $i$, $1 \leq i \leq n$, we can find a subset $Y_i \subseteq Y$ with $\|N[b_i]\|$ elements such that $Y_i \cap Y_j = \emptyset$ for all $i,j \in \{1,\ldots,n\}$ with $i \neq j$. These subsets ensure that $w$ is always placed at the $(n+1)$st position in the second voter group of $V$ below.

**Padding candidates in $Z$:** $Z$ is a set of $(k-1)(n+1)$ candidates such that for each $j$, $1 \leq j \leq k-1$, we can find a subset $Z_j \subseteq Z$ with $n+1$ elements such that $Z_i \cap Z_j = \emptyset$ for all $i,j \in \{1,\ldots,k-1\}$ with $i \neq j$. These subsets ensure that no other candidate besides $c$ and the candidates in $Z_j$ gain any points up to the $(n+2)$nd level in the third voter group of $V$ below.

$V$ is the following collection of $2n+k-1$ voters:

| # | For each … | number of votes | ranking of candidates |
|---|---|---|---|
| 1 | $i \in \{1,\ldots,n\}$ | 1 | $N[b_i]$  $c$  $X_i$  $((B-N[b_i]) \cup (X-X_i) \cup Y \cup Z)$  $w$, |
| 2 | $i \in \{1,\ldots,n\}$ | 1 | $(B-N[b_i])$  $Y_i$  $w$  $(N[b_i] \cup X \cup (Y-Y_i) \cup Z \cup \{c\})$, |
| 3 | $j \in \{1,\ldots,k-1\}$ | 1 | $c$  $Z_j$  $(B \cup X \cup Y \cup (Z-Z_j))$  $w$, |

The relevant scores in $(C,V)$ can be seen in Table 6 below.

| | $c$ | $w$ | $b_j \in B$ |
|---|---|---|---|
| $score^1$ | $k-1$ | $0$ | $\leq n$ |
| $score^{n+1}$ | $\mathbf{n+k-1}$ | $n$ | $n$ |

Table 6: Level $i$ scores in $(C,V)$ for $i \in \{1, n+1\}$ and the candidates in $B \cup \{c,w\}$.

It holds that $n+k-1 > maj(V) > n$. Since candidate $w$ reaches a strict majority only on the last level but $c$ does so no later than on the $(n+1)$st level, $w$ is not a unique Bucklin winner of this election.

We claim that $G$ has a dominating set of size $k$ if and only if $w$ can be made the unique Bucklin winner by deleting at most $k$ voters.

From left to right: Suppose $G$ has a dominating set $B'$ of size $k$. Delete the corresponding voters from the first voter group (i.e., each voter $v_i$ for which $b_i \in B'$). Let $V'$ denote the resulting set of voters and note that $\|V'\| = 2n-1$. Now, in election $(C,V')$ we have on level $n+1$:

- $score^{n+1}_{(C,V')}(b_j) \leq n-1$ for each $b_j \in B$ (from the first and second voter groups; no $b_j$ can have a score of $n$ on level $n+1$, since $B'$ is a dominating set in $G$, so $B = N[B']$, and all voters $v_i$ corresponding to members $b_i$ of $B'$ have been deleted),

- $score^{n+1}_{(C,V')}(c) = (n-k) + (k-1) = n-1$ (from the first and third voter groups),

- $score^{n+1}_{(C,V')}(x_i) = 1$ for each $x_i \in X$ (from the first voter group),

- $score^{n+1}_{(C,V')}(y_i) = 1$ for each $y_i \in Y$ (from the second voter group),

- $score^{n+1}_{(C,V')}(z_i) = 1$ for each $z_i \in Z$ (from the third voter group), and

- $score^{n+1}_{(C,V')}(w) = n$ (from the second voter group).

That is, only candidate $w$ reaches a strict majority on level $n+1$ in $(C,V')$, so $w$ is the unique Bucklin winner of this election.

From right to left: Suppose $w$ can be made the unique Bucklin winner by deleting at most $k$ voters. Let $V'$ be the set of remaining voters. Observe that deleting less than $k$ voters would make it impossible for candidate $w$ to be the unique Bucklin winner of the election. Indeed, if less than $k$ voters are deleted from $V$, the strict majority threshold for the set $V'$ of remaining voters would exceed $n$. However, since $w$ is ranked last place in all votes except the $n$ votes from the second voter group, $w$ would reach a strict majority no earlier than on the last level and thus would not be the unique Bucklin winner of this election. Clearly, $w$ has to win election $(C, V')$ on level $n + 1$. Since $score_{(C,V)}^{n+1}(b_i) = n = score_{(C,V)}^{n+1}(w)$ for all $i$ with $1 \leq i \leq n$, by deleting these $k$ votes from $V$ each $b_i$ has to lose at least one point on the first $n + 1$ levels. Obviously, no voters from the second voter group can be deleted, for otherwise candidate $w$ would not reach the strict majority threshold on level $n + 1$. Similarly, deleting voters from the third voter group does not make any $b_i \in B$ lose any points up to level $n + 1$. So at least part of the deleted voters have to be from the first voter group, let us say we delete $k' \leq k$. Since *every* candidate $b_i \in B$ has to lose at least one point up to level $n + 1$, the $k'$ deleted voters in $V - V'$ correspond to a dominating set in $G$. If $k' < k$, we can delete voters arbitrarily from the first and/or third voter group until the total allowed number of $k$ deleted voters is reached (that is needed to ensure the right majority threshold in the new election).

Note that this polynomial-time reduction is parameterized, as the given parameter $k$ of $k$-DOMINATING SET is the same parameter $k$ that bounds the number of voters that may be deleted in this control problem. ❏

**Corollary 3.19** *Fallback voting is resistant to constructive control by deleting voters, and is parameterized-resistant when this control problem is parameterized by the number of voters deleted.*

The following theorem is stated without proof, since the algorithm for destructive control by adding voters presented in the proof of Theorem 3.16 can easily be adapted to the deleting-voters case.

**Theorem 3.20** *Fallback voting is vulnerable to destructive control by adding voters.*

Again, since Bucklin voting is a special case of fallback voting, we have the following corollary.

**Corollary 3.21** *Bucklin voting is vulnerable to destructive control by adding voters.*

### 3.4.3 Control by Partition of Voters

**Theorem 3.22** *Bucklin voting is resistant to constructive control by partition of voters in both tie-handling models, TE and TP.*

**Proof.** Susceptibility holds by Lemma 3.3. To show NP-hardness we reduce X3C to our control problems. Let $(B, \mathscr{S})$ be an X3C instance with $B = \{b_1, b_2, \ldots, b_{3m}\}$, $m > 1$, and a collection $\mathscr{S} = \{S_1, S_2, \ldots, S_n\}$ of subsets $S_i \subseteq B$ with $\|S_i\| = 3$ for each $i$, $1 \leq i \leq n$. We define the election $(C, V)$, where $C = B \cup \{c, w, x\} \cup D \cup E \cup F \cup G$ is the set of candidates, $w$ is the distinguished candidate, and $D, E, F$, and $G$ are sets of padding candidates (recall Footnote 13).

**Subsets $B_1, B_2, \ldots, B_n$ of $B$:** These are $n$ subsets of $B$ (in general, not disjoint) that are defined such that each candidate in $B$ gains exactly $n$ points in total up to level $3m$ from the first and the second voter group of $V$ below. With $\ell_j = \|\{S_i \in \mathscr{S} \mid b_j \in S_i\}\|$ for each $j$, $1 \leq j \leq 3m$, these subsets are formally defined by $B_i = \{b_j \in B \mid i \leq n - \ell_j\}$ for each $i$, $1 \leq i \leq n$.

**Padding candidates in $D$:** $D$ is a set of $3nm$ candidates such that for each $i$, $1 \le i \le n$, we can find a subset $D_i \subseteq D$ with $3m - \|B_i\|$ elements such that $D_i \cap D_j = \emptyset$ for all $i, j \in \{1, \dots, n\}$ with $i \ne j$. These subsets ensure that $w$ is always placed at position $3m + 1$ in the second voter group of $V$ below.

**Padding candidates in $E$:** $E$ is a set of $(3m-1)(m+1)$ candidates such that for each $k$, $1 \le k \le m+1$, we can find a subset $E_k \subseteq E$ with $3m - 1$ elements such that $E_i \cap E_j = \emptyset$ for all $i, j \in \{1, \dots, m+1\}$ with $i \ne j$. These subsets ensure that no other candidate besides $c$ and $x$ gains more than one point up to the $(3m+1)$st level in the third voter group of $V$ below.

**Padding candidates in $F$:** $F$ is a set of $(3m+1)(m-1)$ candidates such that for each $l$, $1 \le l \le m-1$, we can find a subset $F_l \subseteq F$ with $3m+1$ elements such that $F_i \cap F_j = \emptyset$ for all $i, j \in \{1, \dots, m-1\}$ with $i \ne j$. These subsets ensure that $c$ does not gain any points up to level $3m+1$ in the fourth voter group of $V$ below.

**Padding candidates in $G$:** $G$ is a set of $n(3m-3)$ candidates such that for each $i$, $1 \le i \le n$, we can find a subset $G_i \subseteq G$ with $3m-3$ elements such that $G_i \cap G_j = \emptyset$ for all $i, j \in \{1, \dots, n\}$ with $i \ne j$. These subsets ensure that no other candidate besides $c$ and those in $S_i$ gains more than one point up to level $3m+1$ in the first voter group of $V$ below.

Let $V$ consist of the following $2n + 2m$ voters:

| # | For each … | number of votes | ranking of candidates |
|---|---|---|---|
| 1 | $i \in \{1, \dots, n\}$ | 1 | $c\ \ S_i\ \ G_i\ \ (G - G_i)\ \ F\ \ D\ \ E\ \ (B - S_i)\ \ w\ \ x$ |
| 2 | $i \in \{1, \dots, n\}$ | 1 | $B_i\ \ D_i\ \ w\ \ G\ \ E\ \ (D - D_i)\ \ F\ \ (B - B_i)\ \ c\ \ x$ |
| 3 | $k \in \{1, \dots, m+1\}$ | 1 | $x\ \ c\ \ E_k\ \ F\ \ (E - E_k)\ \ G\ \ D\ \ B\ \ w$ |
| 4 | $l \in \{1, \dots, m-1\}$ | 1 | $F_l\ \ c\ \ (F - F_l)\ \ G\ \ D\ \ E\ \ B\ \ w\ \ x$ |

The strict majority threshold is reached with $n + m + 1$ points and Table 7 shows the scores on the relevant levels in election $(C, V)$.

| | $c$ | $b_j$ | $w$ | $x$ |
|---|---|---|---|---|
| $score^1$ | $n$ | $\le n$ | $0$ | $m+1$ |
| $score^2$ | $\mathbf{n+m+1}$ | $\le n$ | $0$ | $m+1$ |
| $score^{3m}$ | $n+m+1$ | $n$ | $0$ | $m+1$ |
| $score^{3m+1}$ | $n+m+1$ | $n$ | $n$ | $m+1$ |

Table 7: Level $i$ scores in $(C, V)$ for $i \in \{1, 2, 3m, 3m+1\}$ and the candidates in $B \cup \{c, w, x\}$.

Clearly, candidate $c$ is the unique level 2 BV winner in $(C, V)$ with a level 2 score of $n + m + 1$.

We claim that $\mathscr{S}$ has an exact cover $\mathscr{S}'$ for $B$ if and only if $w$ can be made the unique BV winner of the resulting election by partition of voters (regardless of the tie-handling model used).

From left to right: Suppose $\mathscr{S}$ has an exact cover $\mathscr{S}'$ for $B$. Partition $V$ as follows. Let $V_1$ consist of:

- the $m$ voters of the first group that correspond to the exact cover (i.e., those $m$ voters of the form $c\ \ S_i\ \ G_i\ \ (G - G_i)\ \ F\ \ D\ \ E\ \ (B - S_i)\ \ w\ \ x$ for which $S_i \in \mathscr{S}'$) and

- the $m+1$ voters of the third group (i.e., all voters of the form $x \ c \ E_k \ F \ (E-E_k) \ G \ D \ B \ w$).

Let $V_2 = V - V_1$. Table 8 shows the relevant scores in the two subelections.

| | $(C,V_1)$ | | | $(C,V_2)$ | | |
|---|---|---|---|---|---|---|
| | $c$ | $b_j$ | $x$ | $c$ | $b_j$ | $w$ |
| $score^1$ | $m$ | $0$ | $\mathbf{m+1}$ | $n-m$ | $\leq n-1$ | $0$ |
| $score^2$ | $2m+1$ | $\leq 1$ | $m+1$ | $n-m$ | $\leq n-1$ | $0$ |
| $score^{3m}$ | $2m+1$ | $1$ | $m+1$ | $n-m$ | $n-1$ | $0$ |
| $score^{3m+1}$ | $2m+1$ | $1$ | $m+1$ | $n-m$ | $n-1$ | $\mathbf{n}$ |

Table 8: Level $i$ scores in $(C,V_1)$ and $(C,V_2)$ for $i \in \{1,2,3m,3m+1\}$ and the candidates in $B \cup \{c,w,x\}$.

In subelection $(C,V_1)$, candidate $x$ is the unique level 1 BV winner. In subelection $(C,V_2)$, candidate $w$ is the first candidate who has a strict majority and moves on to the final round of the election. Thus there are $w$ and $x$ in the final run-off, which $w$ wins with a strict majority on the first level as can be seen in Table 9 presenting the scores of $x$ and $w$ in the final election.

| | $x$ | $w$ |
|---|---|---|
| $score^1$ | $m+1$ | $\mathbf{2n+m-1}$ |
| $score^2$ | $2n+2m$ | $2n+2m$ |

Table 9: Level $i$ scores of $w$ and $x$ in the final election $(\{w,x\},V)$ for $i \in \{1,2\}$.

Since both subelections, $(C,V_1)$ and $(C,V_2)$, have unique BV winners, candidate $w$ can be made the unique BV winner by partition of voters, regardless of the tie-handling model used.

From right to left: Suppose that $w$ can be made the unique BV winner by exerting control by partition of voters (for concreteness, say in TP). Let $(V_1,V_2)$ be such a successful partition. Since $w$ wins the resulting two-stage election, $w$ has to win at least one of the subelections (say, $w$ wins $(C,V_2)$). If candidate $c$ participates in the final round, he or she wins the election with a strict majority no later than on the second level, no matter which other candidates move to the final election. That means that in both subelections, $(C,V_1)$ and $(C,V_2)$, $c$ must not be a BV winner. Only in the second voter group candidate $w$ (who has to be a BV winner in $(C,V_2)$) gets points earlier than on the second-to-last level. So $w$ has to be a level $3m+1$ BV winner in $(C,V_2)$ via votes from the second voter group in $V_2$. As $c$ scores already on the first two levels in voter groups 1 and 3, only $x$ and the candidates in $B$ can prevent $c$ from winning in $(C,V_1)$. However, since voters from the second voter group have to be in $V_2$ (as stated above), in subelection $(C,V_1)$ only candidate $x$ can prevent $c$ from moving forward to the final round. Since $x$ is always placed behind $c$ in all votes except those votes from the third voter group, $x$ has to be a level 1 BV winner in $(C,V_1)$. In $(C,V_2)$ candidate $w$ gains all the points on exactly the $(3m+1)$st level, whereas the other candidates scoring more than one point up to this level receive their points on either earlier or later levels, so no candidate can tie with $w$ on the $(3m+1)$st level and $w$ is the unique level $3m+1$ BV winner in $(C,V_2)$. As both subelections, $(C,V_1)$ and $(C,V_2)$, have unique BV winners other than $c$, the construction works in model TE as well.

It remains to show that $\mathscr{S}$ has an exact cover $\mathscr{S}'$ for $B$. Since $w$ has to win $(C,V_2)$ with the votes from the second voter group, not all voters from the first voter group can be in $V_2$ (otherwise $c$ would have $n$ points already on the first level). On the other hand, there can be at most $m$ voters from the first voter group in $V_1$ because otherwise $x$ would not be a level 1 BV winner in $(C,V_1)$. To ensure that no candidate contained in $B$ has the same score as $w$, namely $n$ points, and gets these points on an earlier level than $w$ in $(C,V_2)$, there have to be exactly $m$ voters from the first group in $V_1$ and these voters correspond to an exact cover for $B$. $\qquad\Box$

Since Bucklin voting is a special case of fallback voting, fallback voting inherits the NP-hardness lower bounds from Bucklin voting stated in Theorem 3.22.

**Corollary 3.23** *Fallback voting is resistant to constructive control by partition of voters in model TE and model TP.*

We now turn to destructive control by partition of voters in model TE where we show resistance for Bucklin and fallback voting via a reduction from the dominating set problem that has been defined in Section 2.5. The following construction will be used in the proof of Theorem 3.26.

**Construction 3.24** *Let $((B,A),k)$ be a given instance of DOMINATING SET with $B = \{b_1,b_2,\ldots,b_n\}$ and $n \geq 1$. Define the election $(C,V)$ with candidate set*

$$C = B \cup D \cup E \cup F \cup G \cup \{c,u,v,w,x,y\},$$

*where $c$ is the distinguished candidate, $D$, $E$, $F$, $G$, and $\{u,v\}$ are sets of padding candidates (recall Footnote 13), and $y$ is a "partition-enforcing" candidate (see below).*

**Padding candidates in $D$:** *$D$ is a set of $(k-1)(n+4)$ candidates such that for each $j$, $1 \leq j \leq k-1$, we can find a subset $D_j \subseteq D$ with $n+4$ elements such that $D_i \cap D_j = \emptyset$ for all $i,j \in \{1,\ldots,k-1\}$ with $i \neq j$. These subsets ensure that no other candidate besides $x$ gains more than one point up to level $n+5$ in the third voter group of $V$ below.*

**Padding candidates in $E$:** *$E$ is a set of $2(k+n)$ candidates such that for each $l$, $1 \leq l \leq m+1$, we can find a subset $E_l \subseteq E$ with two elements and it holds that $E_i \cap E_j = \emptyset$ for all $i,j \in \{1,\ldots,k+n\}$ with $i \neq j$. These subsets ensure that $x$ and $y$ do not gain any points up to the fourth level in the fourth voter group of $V$ below.*

**Padding candidates in $F$:** *$F$ is a set of $3n$ candidates such that for each $i$, $1 \leq i \leq n$, we can find a subset $F_i \subseteq F$ with three elements such that $F_i \cap F_j = \emptyset$ for all $i,j \in \{1,\ldots,n\}$ with $i \neq j$. These subsets ensure that the candidates in $B$ do not gain any points up to the fourth level in the first voter group of $V$ below.*

**Padding candidates in $G$:** *$G$ is a set of $n^2$ candidates such that for each $i$, $1 \leq i \leq n$, we can find a subset $G_i \subseteq G$ with $\|N[b_i]\|$ elements such that $G_i \cap G_j = \emptyset$ for all $i,j \in \{1,\ldots,n\}$ with $i \neq j$. These subsets ensure that $w$ does not gain any points up to level $n+5$ in the first voter group of $V$ below.*

**Padding candidates *u* and *v*:** *These two candidates ensure that the other padding candidates are not among the top $n + 5$ positions in the second voter group of $V$ below.*

**Partition-enforcing candidate *y*:** *This candidate ensures that the voter from the second voter group of $V$ below has to be in the subelection candidate $w$ wins to finally beat $c$ in the final election.*

*$V$ consists of the following $2k + 2n$ votes that can be arranged in four groups:*

| # | For each . . . | number of votes | ranking of candidates |
|---|---|---|---|
| 1 | $i \in \{1, \ldots, n\}$ | 1 | $F_i$ $(B - \mathrm{N}[b_i])$ $H_i$ $y$ $w \cdots$ |
| | | | $\cdots (\mathrm{N}[b_i] \cup D \cup E \cup (F - F_i) \cup (H - H_i))$ $u$ $v$ $c$ $x$ |
| 2 | | 1 | $x$ $w$ $c$ $B$ $u$ $v$ $(D \cup E \cup F \cup H)$ $y$ |
| 3 | $j \in \{1, \ldots, k-1\}$ | 1 | $x$ $D_j$ $(B \cup (D - D_j) \cup E \cup F \cup H)$ $u$ $v$ $y$ $w$ $c$ |
| 4 | $l \in \{1, \ldots, k+n\}$ | 1 | $c$ $E_l$ $x$ $y$ $(B \cup D \cup (E - E_l) \cup F \cup H)$ $u$ $v$ $w$ |

Table 10 shows the scores of $c$, $w$, and $x$ on the first three levels. None of the other candidates scores more than one point up to the third level. Note that $c$ reaches a strict majority on this level and thus is the unique level 3 BV winner in this election.

| | $c$ | $w$ | $x$ |
|---|---|---|---|
| $score^1$ | $k + n$ | $0$ | $k$ |
| $score^2$ | $k + n$ | $1$ | $k$ |
| $score^3$ | $\mathbf{k + n + 1}$ | $1$ | $k$ |

Table 10: Level $i$ scores of $c$, $w$, and $x$ in $(C, V)$ for $i \in \{1, 2, 3\}$.

**Lemma 3.25** *In the election $(C, V)$ from Construction 3.24, for every partition of $V$ into $V_1$ and $V_2$, candidate $c$ is the unique BV winner of at least one of the subelections, $(C, V_1)$ and $(C, V_2)$.*

**Proof.** For a contradiction, we assume that in both subelections, $(C, V_1)$ and $(C, V_2)$, candidate $c$ is not a unique BV winner. Table 10 shows that half of the voters in $V$ place $c$ already on the first level. So the following must hold:

- Both $\|V_i\|$ must be even numbers for $i \in \{1, 2\}$ and

- $score^1_{(C, V_i)}(c) = \|V_i\|/2$ for $i \in \{1, 2\}$.

Because of the voter in the second voter group, candidate $c$ will get a strict majority on the third level in one of the subelections, let us say in $(C, V_1)$. So there has to be a candidate beating or tieing with candidate $c$ on the second or third level in $(C, V_1)$. The candidates in $B$, $D$, $E$, $F$, and $H$ and the candidates $u$, $v$, $w$, and $y$ do not score more than one point up to the third level. Thus only candidate $x$ can possibly beat or tie with $c$ on the second or third level in $(C, V_1)$. However, since $x$ does not score more than $k$ points in total until the fourth level, $c$ is the unique level 3 BV winner in $(C, V_1)$, a contradiction. It follows that $c$ is a unique BV winner of at least one of the subelections. ❑

**Theorem 3.26** *Bucklin voting is resistant to destructive control by partition of voters in model TE, and is parameterized-resistant when this control problem is parameterized by the size of the smaller partition.*

**Proof.** Susceptibility follows from Lemma 3.3. To prove NP-hardness, we again provide a reduction from the NP-complete problem DOMINATING SET that has been defined in Section 2.5. Given a DOMINATING SET instance $((B,A),k)$, construct a Bucklin election $(C,V)$ according to Construction 3.24.

We claim that $G = (B,A)$ has a dominating set $B'$ of size $k$ if and only if candidate $c$ can be prevented from being a unique BV winner by partition of voters in model TE.

From left to right: Let $B'$ be a dominating set for $G$ of size $k$. Partition $V$ into $V_1$ and $V_2$ as follows. Let $V_1$ consist of the following $2k$ voters:

- The voters of the first voter group corresponding to the dominating set, i.e., the $k$ voters of the form:

$$F_i \ (B - \mathrm{N}[b_i]) \ H_i \ y \ w \ (\mathrm{N}[b_i] \cup D \cup E \cup (F - F_i) \cup (H - H_i)) \ u \ v \ c \ x$$

  for those $i$ for which $b_i \in B'$,

- the one voter from the second group:

$$x \ w \ c \ B \ u \ v \ (D \cup E \cup F \cup H) \ y,$$

  and

- the entire third voter group, i.e., the $k-1$ voters of the form:

$$x \ D_j \ (B \cup (D - D_j) \cup E \cup F \cup H \cup) \ u \ v \ y \ w \ c,$$

  where $1 \le j \le k-1$.

Let $V_2 = V - V_1$. Note that the strict majority threshold in $V_1$ is $maj(V_1) = k + 1$. Again, since the candidates in $D$, $E$, $F$, and $H$ do not score more than one point up to level $n + 5$, their level $n + 5$ scores are not shown in Table 11. The level $n + 5$ scores of the remaining candidates are shown in this table. Note that $w$ reaches a strict majority of $k + 1$ on this level (and no other candidate reaches a strict majority on this or an earlier level). Hence, $w$ is the unique level $n + 5$ BV winner in subelection $(C, V_1)$ and thus participates in the final round.

| | $c$ | $w$ | $x$ | $y$ | $b_i \in B$ |
|---|---|---|---|---|---|
| $score^{n+5}$ | 1 | **$k+1$** | $k$ | $k$ | $\le k$ |

Table 11: Level $n+5$ scores in $(C, V_1)$.

From Lemma 3.25 it follows that candidate $c$ is the unique winner in subelection $(C, V_2)$. So the final-stage election is $(\{c, w\}, V)$ and we have the following scores on the first two levels:

$$score^1_{(\{c,w\},V)}(c) = score^1_{(\{c,w\},V)}(w) \ = \ k + n,$$
$$score^2_{(\{c,w\},V)}(c) = score^2_{(\{c,w\},V)}(w) \ = \ 2k + 2n.$$

Since none of $c$ and $w$ have a strict majority on the first level, both candidates are level 2 BV winners in this two-candidate final-stage election. Hence, $c$ has been prevented from being a unique BV winner by partition of voters in model TE.

From right to left: Assume that $c$ can be prevented from being a unique BV winner by partition of voters in model TE. From Lemma 3.25 we know that candidate $c$ must participate in the final-stage election. Since we are in model TE, at most two candidates participate in the final run-off. To prevent $c$ from being a unique BV winner of the final election, there must be another finalist and this other candidate has to beat or tie with $c$. Since $w$ is the only candidate that can beat or tie with $c$ in a two-candidate election, $w$ has to move on to the final round to run against $c$. Let us say that $c$ is the unique winner of subelection $(C, V_2)$ and $w$ is the unique winner of subelection $(C, V_1)$. For $w$ to be the unique winner of subelection $(C, V_1)$, $V_1$ has to contain voters from the first voter group and $w$ can win only on the $(n+5)$th level. In particular, $x$ is placed before $w$ in all voter groups except the first, so $w$ can win in $(C, V_1)$ only via voters from the first voter group participating in $(C, V_1)$. Moreover, since $w$ is placed in the last or second-to-last position in all voters from the third and fourth groups, and since there is only one voter in the second group, $w$ can win only on the $(n+5)$th level (which is $w$'s position in the votes from the first voter group).

Let $I \subseteq \{1, \ldots, n\}$ be the set of indices $i$ such that first-group voter

$$F_i \ (B - \mathrm{N}[b_i]) \ H_i \ y \ w \ (\mathrm{N}[b_i] \cup D \cup E \cup (F - F_i) \cup (H - H_i)) \ u \ v \ c \ x$$

belongs to $V_1$. Let $\ell = \|I\|$. Since $w$ is the unique level $n+5$ BV winner of $(C, V_1)$ but $y$ is placed before $w$ in every vote in the first group, the one voter from the second group (which is the only voter who prefers $w$ to $y$) must belong to $V_1$. Thus we know that

$$score^{n+5}_{(C,V_1)}(w) = \ell + 1 \quad \text{and} \quad score^{n+4}_{(C,V_1)}(y) = score^{n+5}_{(C,V_1)}(y) = \ell.$$

For the candidates in $B$, we have

$$score^{n+4}_{(C,V_1)}(b_j) = score^{n+5}_{(C,V_1)}(b_j) = 1 + \|\{b_i \mid i \in I \text{ and } b_j \notin \mathrm{N}[b_i]\}\|,$$

since each $b_j$ scores one point up to the $(n+4)$th level from the voter in the second group and one point from the first group for every $b_i$ with $i \in I$ such that $b_j \notin \mathrm{N}[b_i]$ in graph $G$. Again, since $w$ is the unique level $n+5$ BV winner of $(C, V_1)$, no $b_j \in B$ can score a point in *each* of the $\ell$ votes from the first voter group that belong to $V_2$. This implies that for each $b_j \in B$ there has to be at least one $b_i$ with $i \in I$ that is adjacent to $b_j$ in $G$. Thus, the set $B'$ of candidates $b_i$ with $i \in I$ corresponds to a dominating set in $G$.

Recall that $score^{n+5}_{(C,V_1)}(w) = \ell + 1$ and $score^{n+4}_{(C,V_1)}(y) = \ell$. Note also that $score^{n+4}_{(C,V_1)}(b_j) \le \ell$ for $1 \le j \le n$. Since $w$ needs a strict majority to be a BV winner in $(C, V_1)$, it must hold that $maj(V_1) \le \ell + 1$. Since $y$ and the $b_j \in B$ have a score of $\ell$ already one level earlier than $w$, it must hold that $maj(V_1) = \ell + 1$, which implies $\|V_2\| = 2\ell$ or $\|V_2\| = 2\ell + 1$. To ensure this cardinality of $V_1$, other votes have to be added. Since $y$ must not gain additional points from these votes up to the $(n+5)$th level, they cannot come from the fourth voter group. The remaining votes from the third voter group total up to $k-1$. Thus, since $w$ is the unique BV winner in subelection $(C, V_1)$, it must hold that $\ell \le k$. So $\|B'\| = \ell \le k$ and this means that there exists a dominating set of size at most $k$.

Observe that reducing from the W[2]-complete parameterized problem $k$-Dominating Set with the same construction and the new parameter $k' = 2k$ (the size of the smaller partition), we can prove that destructive control by partition of voters in model TE in Bucklin voting is W[2]-hard. ❑

Resistance (and parameterized resistance) for fallback voting to this control type now follows immediately.

**Corollary 3.27** *Fallback voting is resistant to destructive control by partition of voters in model TE, and is parameterized-resistant when this control problem is parameterized by the size of the smaller partition.*

The following construction will be used to handle the case of destructive control by partition of voters in model TP (see Theorem 3.30 below) for fallback voting. Construction 3.28 starts from an instance of the NP-complete problem Restricted Hitting Set defined in Section 2.5.

**Construction 3.28** *Let $(B, \mathscr{S}, k)$ be a given instance of* Restricted Hitting Set, *where $B = \{b_1, b_2, \ldots, b_m\}$ is a set, $\mathscr{S} = \{S_1, S_2, \ldots, S_n\}$ is a collection of nonempty subsets $S_i \subseteq B$ such that $n > m$, and $k$ is an integer with $1 < k < m$.*

*Define the election $(C, V)$, where*

$$C = B \cup D \cup E \cup \{c, w\}$$

*is the candidate set with $D = \{d_1, \ldots, d_{2(m+1)}\}$ and $E = \{e_1, \ldots, e_{2(m-1)}\}$. Note that the candidates contained in $D$ and $E$ are padding candidates (recall Footnote 13). In particular, the candidates in $D$ ensure that $w$ is always placed at the third position in the votes of the fourth voter group of $V$ below. The collection of voters $V$ consists of the following $2n(k+1) + 4m + 2mk$ voters:*

| # | For each … | number of voters | ranking of approved candidates |
|---|------------|------------------|-------------------------------|
| 1 | $i \in \{1, \ldots, n\}$ | $k + 1$ | $w$ $S_i$ $c$ |
| 2 | $j \in \{1, \ldots, m\}$ | 1 | $c$ $b_j$ $w$ |
| 3 | $j \in \{1, \ldots, m\}$ | $k - 1$ | $b_j$ |
| 4 | $p \in \{1, \ldots, m+1\}$ | 1 | $d_{2(p-1)+1}$ $d_{2p}$ $w$ |
| 5 | $r \in \{1, \ldots, 2(m-1)\}$ | 1 | $e_r$ |
| 6 | | $n(k+1) + m - k + 1$ | $c$ |
| 7 | | $mk + k - 1$ | $c$ $w$ |
| 8 | | 1 | $w$ $c$ |

The strict majority threshold for $V$ is $maj(V) = n(k+1) + 2m + mk + 1$. In election $(C, V)$, only the two candidates $c$ and $w$ reach a strict majority, $w$ on the third level and $c$ on the second level (see Table 12). Thus $c$ is the unique level 2 FV winner of election $(C, V)$.

Lemma 3.29 will be used in the proof of Theorem 3.30.

**Lemma 3.29** *In the election $(C, V)$ from Construction 3.28, for every partition of $V$ into $V_1$ and $V_2$, candidate $c$ is an FV winner of $(C, V_1)$ or $(C, V_2)$.*

| | $c$ | $w$ | $b_j \in B$ | $d_p \in D$ | $e_r \in E$ |
|---|---|---|---|---|---|
| score$^1$ | $n(k+1)+2m+mk$ | $n(k+1)+1$ | $k-1$ | $\leq 1$ | $1$ |
| score$^2$ | $\mathbf{n(k+1)+2m+mk+1}$ | $n(k+1)+mk+k$ | $\leq k+n(k+1)$ | $1$ | $1$ |
| score$^3$ | $\leq 2n(k+1)+2m+mk+1$ | $n(k+1)+2m+mk+k+1$ | $\leq k+n(k+1)$ | $1$ | $1$ |
| score$^{m+2}$ | $2n(k+1)+2m+mk+1$ | $n(k+1)+2m+mk+k+1$ | $\leq k+n(k+1)$ | $1$ | $1$ |

Table 12: Level $i$ scores for $i \in \{1,2,m+2\}$ in the election $(C,V)$ from Construction 3.28.

**Proof.** For a contradiction, suppose that in both subelections, $(C,V_1)$ and $(C,V_2)$, candidate $c$ is not an FV winner. Since $score^1_{(C,V)}(c) = \|V\|/2$, the two subelections must satisfy that

- both $\|V_1\|$ and $\|V_2\|$ are even numbers, and

- $score^1_{(C,V_1)}(c) = \|V_1\|/2$ and $score^1_{(C,V_2)}(c) = \|V_2\|/2$.

Otherwise, $c$ would have a strict majority already on the first level in one of the subelections and would win that subelection. For each $i \in \{1,2\}$, $c$ already on the first level has only one point less than the strict majority threshold $maj(V_i)$ in subelection $(C,V_i)$, and $c$ will get a strict majority in $(C,V_i)$ no later than on the $(m+2)$nd level. Thus, for both $i=1$ and $i=2$, there must be candidates whose level $m+2$ scores in $(C,V_i)$ are higher than the level $m+2$ score of $c$ in $(C,V_i)$. Table 12 shows the level $m+2$ scores of all candidates in $(C,V)$. Only $w$ and some $b_j \in B$ have a chance to beat $c$ on that level in $(C,V_i)$, $i \in \{1,2\}$.

Suppose that $c$ is defeated in both subelections by two distinct candidates from $B$ (say, $b_x$ defeats $c$ in $(C,V_1)$ and $b_y$ defeats $c$ in $(C,V_2)$). Thus the following must hold:[15]

$$score^{m+2}_{(C,V_1)}(b_x) + score^{m+2}_{(C,V_2)}(b_y) \geq score^{m+2}_{(C,V)}(c) + 2$$
$$2n(k+1) + 2k - n(k+1) \geq 2n(k+1) + mk + 2m + 3$$
$$2k \geq n(k+1) + mk + 2m + 3,$$

which by our basic assumption $m > k > 1$ implies the following contradiction:

$$0 \geq n(k+1) + (m-2)k + 2m + 3 > n(k+1) + (k-2)k + 2k + 3 = n(k+1) + k^2 + 3 > 0.$$

Thus the only possibility for $c$ to not win any of the two subelections is that $c$ is defeated in one subelection, say $(C,V_1)$, by a candidate from $B$, say $b_x$, and in the other subelection, $(C,V_2)$, by candidate $w$. Then it must hold that:[15]

$$score^{m+2}_{(C,V_1)}(b_x) + score^{m+2}_{(C,V_2)}(w) \geq score^{m+2}_{(C,V)}(c) + 2$$
$$2n(k+1) + 2k + 2m + mk + 1 - n(k+1) - 1 \geq 2n(k+1) + mk + 2m + 3$$
$$2k \geq n(k+1) + 3.$$

Since $n > 1$, this cannot hold, so $c$ must be an FV winner in one of the two subelections. ❑

---

[15]For the left-hand sides of the inequalities, note that each vote occurs in only one of the two subelections. To avoid double-counting those votes that give points to both candidates, we first sum up the overall number of points each candidate scores and then substract the double-counted points.

**Theorem 3.30** *Fallback voting is resistant to destructive control by partition of voters in model TP.*

**Proof.** Susceptibility holds by Corollary 3.4. To prove NP-hardness, we reduce RESTRICTED HITTING SET to our control problem. Consider the election $(C,V)$ constructed according to Construction 3.28 from a given RESTRICTED HITTING SET instance $(B, \mathscr{S}, k)$, where $B = \{b_1, \dots, b_m\}$ is a set, $\mathscr{S} = \{S_1, \dots, S_n\}$ is a collection of nonempty subsets $S_i \subseteq B$, and $k$ is an integer with $1 < k < m < n$.

We claim that $\mathscr{S}$ has a hitting set $B' \subseteq B$ of size $k$ if and only if $c$ can be prevented from being a unique FV winner by partition of voters in model TP.

From left to right: Suppose, $B' \subseteq B$ is a hitting set of size $k$ for $\mathscr{S}$. Partition $V$ into $V_1$ and $V_2$ the following way. Let $V_1$ consist of those voters of the second group where $b_j \in B'$ and of those voters of the third group where $b_j \in B'$. Let $V_2 = V - V_1$. In $(C, V_1)$, no candidate reaches a strict majority (see Table 13), where $maj(V_1) = \lfloor k^2/2 \rfloor + 1$, and candidates $c$, $w$, and each $b_j \in B'$ win the election with an approval score of $k$.

| | $c$ | $w$ | $b_j \in B'$ | $b_j \notin B'$ |
|---|---|---|---|---|
| score[1] | $k$ | $0$ | $k-1$ | $0$ |
| score[2] | $k$ | $0$ | $k$ | $0$ |
| score[3] | **k** | **k** | **k** | $0$ |

Table 13: Level $i$ scores in $(C, V_1)$ for $i \in \{1, 2, 3\}$ and the candidates in $B \cup \{c, w\}$.

| | $c$ | $w$ | $b_j \notin B'$ | $b_j \in B'$ |
|---|---|---|---|---|
| score[1] | $n(k+1) + 2m - k + mk$ | $n(k+1) + 1$ | $k-1$ | $0$ |
| score[2] | $n(k+1) + 2m - k + mk + 1$ | $n(k+1) + mk + k$ | $\leq k + n(k+1)$ | $\leq n(k+1)$ |
| score[3] | $\geq n(k+1) + 2m - k + mk + 1$ | $n(k+1) + mk + 2m + 1$ | $\leq k + n(k+1)$ | $\leq n(k+1)$ |

Table 14: Level $i$ scores in $(C, V_2)$ for $i \in \{1, 2, 3\}$ and the candidates in $B \cup \{c, w\}$.

The level $i$ scores in election $(C, V_2)$ for $i \in \{1, 2, 3\}$ and the candidates in $B \cup \{c, w\}$ are shown in Table 14. Since in $(C, V_2)$ no candidate from $B$ wins, the candidates participating in the final round are $B' \cup \{c, w\}$. The scores in the final election $(B' \cup \{c, w\}, V)$ can be seen in Table 15. Since candidates $c$ and $w$ with the same level 2 scores are both level 2 FV winners, candidate $c$ has been prevented from being a unique FV winner by partition of voters in model TP.

| | $c$ | $w$ | $b_j \in B'$ |
|---|---|---|---|
| score[1] | $n(k+1) + 2m + mk$ | $n(k+1) + m + 2$ | $k-1$ |
| score[2] | **n(k+1) + 2m + mk + 1** | **n(k+1) + 2m + mk + 1** | $\leq k + n(k+1)$ |

Table 15: Level $i$ scores in the final-stage election $(B' \cup \{c, w\}, V)$ for $i \in \{1, 2\}$.

From right to left: Suppose candidate $c$ can be prevented from being a unique FV winner by partition of voters in model TP. From Lemma 3.29 it follows that candidate $c$ participates in the final round. Since $c$ has a strict majority of approvals, $c$ has to be tied with or lose against another candidate by a strict majority at some level. Only candidate $w$ has a strict majority of approvals, so $w$ has to tie or beat $c$ at some level in the final round. Because of the low scores of the candidates in $D$ and $E$ we may assume that only candidates from $B$ are participating in the final round besides $c$ and $w$. Let $B' \subseteq B$ be the set of candidates who also participate in the final round. Let $\ell$ be the number of sets in $\mathscr{S}$ not hit by $B'$. As $w$ cannot reach a strict majority of approvals on the first level, we consider the level 2 scores of $c$ and $w$:

$$
\begin{aligned}
score^2_{(B' \cup \{c,w\}, V)}(c) &= n(k+1) + 2m + mk + 1 + \ell(k+1), \\
score^2_{(B' \cup \{c,w\}, V)}(w) &= n(k+1) + 2m + mk + k - \|B'\| + 1.
\end{aligned}
$$

Since $c$ has a strict majority already on the second level, $w$ must tie or beat $c$ on this level, so the following must hold:

$$
\begin{aligned}
score^2_{(B' \cup \{c,w\}, V)}(c) - score^2_{(B' \cup \{c,w\}, V)}(w) &\leq 0 \\
n(k+1) + 2m + mk + 1 + \ell(k+1) - n(k+1) - 2m - mk - k + \|B'\| - 1 &\leq 0 \\
\|B'\| - k + \ell(k+1) &\leq 0.
\end{aligned}
$$

This is possible only if $\ell = 0$ (i.e., all sets in $\mathscr{S}$ are hit by $B'$), which implies $\|B'\| \leq k$. Thus $\mathscr{S}$ has a hitting set of size at most $k$. $\qquad\qquad\Box$

## 4   Conclusions and Open Questions

We have shown that, among natural election systems with a polynomial-time winner problem, fallback voting displays the broadest control resistance currently known to hold. We have also shown that Bucklin voting behaves almost as good (possibly even as good) as fallback voting in terms of control resistance. In particular, both voting systems are resistant to all standard types of candidate control and to all standard types of constructive control. In total, fallback voting has 20 resistances and two vulnerabilities and Bucklin voting possesses at least 19 (possibly even 20) resistances and has at least two (and can have no more than three) vulnerabilities. One case remains open: destructive control by partition of voters in the tie-handling model "ties promote" for Bucklin voting. For comparison, recall from Table 1 that approval voting is vulnerable to destructive control by partition of voters both in model TE and in model TP and that SP-AV is vulnerable to this control type in model TE but resistant in model TP.

With this paper the research program that was started by Bartholdi et al. [BTT92] two decades ago may now be viewed as "essentially completed."

We also strengthened some of our resistance (i.e., NP-hardness) results for (unparameterized) control problems by showing that their parameterized variants for natural parameterizations are even W[2]-hard. It remains open whether or not these problems belong to W[2]. Another interesting task for future research along these lines is to find natural parameterizations for the other problems (including some problems modeling control by partition of either candidates or voters) and to study their parameterized complexity.

From a theoretical point of view, it would also be interesting and challenging to do a typical-case analysis for control problems as has been done for manipulation (see, e.g., [CSL07, PR07, FKN08, XC08a, XC08b] and the survey [RS12c]).

## Acknowledgments

We thank the anonymous AAMAS-2011, CATS-2010, COMSOC-2010, and JAIR referees for their very helpful comments and suggestions.